\documentclass[12pt]{article}
\usepackage{amssymb}
\usepackage{graphicx}
\begin{document}
\title{On polynomials interpolating between the stationary state of a $O(n)$ model and a Q.H.E. ground state}
\author {  M. Kasatani$^1$ and V. Pasquier$^2$}
\date{}
\maketitle \hskip-6mm {$^1$  Department of Mathematics, Kyoto
university, Kyoto 606-8502, Japan.}\\ { $^2$ Service de Physique
Th\'eorique, C.E.A/ Saclay, 91191 Gif-sur-Yvette, France.}

\begin{abstract}

We obtain a family of polynomials defined by vanishing conditions
and associated to tangles. We study more specifically the case
where they are related to a $O(n)$ loop model. We conjecture that
their specializations at $z_i=1$ are {\it positive} in $n$. At
$n=1$, they coincide with the the Razumov-Stroganov integers
counting alternating sign matrices.

We derive the CFT modular invariant partition functions labelled
by Coxeter-Dynkin diagrams using the representation theory of the
affine Hecke algebras.
\end{abstract}
\maketitle

\section{Introduction}

Much progress has recently been made in the study of the ground
state of the $XXZ$ spin chain Hamiltonian when the anisotropy
parameter is equal to ${1/2}$ \cite{razumov1}. The Hamiltonian is
closely related to a stochastic Hamiltonian of a $O(1)$ fully
packed loop model \cite{pierce,batchelor}.

The components of the $O(1)$ model Hamiltonian stationary state
can be normalized to be positive integers, and it is conjectured
(the Razumov-Stroganov (R-S) conjecture \cite{razumov2}) that
these integers are in bijective correspondence with the states of
a square-ice-model with domain wall boundary conditions, or
equivalently with certain classes of alternating-sign matrices.

In previous works \cite{kasatani}\cite{pas1}, we have introduced
the polynomials discussed in this paper. In \cite{kasatani}, they
were discovered through the study of some representations of the
double-affine Hecke algebra \cite{cherednick2}. In \cite{pas1},
they were obtained by deforming the $O(1)$ transfer matrix
eigenstates of Di Francesco and Zinn-Justin \cite{Pdf} in order to
generalize the the R-S conjecture to the $O(n)$ case.

The aim of this paper is to further study these polynomials:

\begin{itemize}
\item We construct a family of polynomials which transform
linearly under the braid group. We single out a basis in
correspondence with the flat tangles or patterns.

 \item We define a deformation of the
R-S integers by evaluating the basis polynomials at $1$. We
observe they enjoy positivity properties suggesting that they  may
coincide with a weighted enumeration sum of objects related to
alternating sign matrices.
\end{itemize}
\bigskip

The polynomials can be defined by the vanishing conditions which
they obey when several variables come close to each other. These
vanishing conditions, called the wheel conditions \cite{feigin},
have been classified in \cite{kasatani} in relation with the
representation theory of affine Hecke algebras. The number of
variables involved in the wheel condition is $k+1$ and the wheel
condition depends on another parameter $r$ related to the degree
with which the polynomial vanishes when the points are put
together. In an orthogonal basis, the polynomials are
non-symmetric Macdonald polynomials at specialized values of their
parameters $t^{k+1}q^{r-1}=1$.\footnote{ The non-symmetrical
Macdonald polynomials  were first introduced in the context of the
Calogero-Sutherland models by diagonalizing the affine Hecke
generators $\bar y_i$ (constructed in appendix
\ref{section.diag-yi}) in the published version of \cite{BGHP}.
They have been  extensively used by Cherednick
\cite{cherednick2}.}

Another way to obtain them, giving rise to a different basis, is
by deforming the $O(1)$ model transfer matrix ground state. When
$k=2$, the polynomial representation is dual to a representation
of the Temperley and Lieb (T.L.) algebra of a loop model (defined
on the disc) parameterized by $n=-2\cos {(\theta)}$ where $n$ is
the fugacity of the loops. The polynomials basis we consider here
is dual to the loop model basis. Its specialization at $n=1$
coincides with the components of the $O(1)$ model transfer matrix
ground state.

The basis polynomials are obtained from the action of difference
operators \cite{lasc2} on a generating state which in the $r=2$
case is simply a product of q-deformed Vandermonde determinants.
They can also be defined as the Kazhdan-Lusztig (K-L) basis
\cite{KL}\cite{lasc1} of this representation.\footnote{The
relation with the Kazhdan-Lusztig basis has been explained to us
by A. Lascoux.}

At $\theta=0$, the wheel conditions are precisely the constraint
imposed by the interactions on the Q.H.E. wave functions and the
generating polynomial of the basis coincides with a wave functions
of the Quantum Hall Effect (Q.H.E) \cite{read}. In this context,
the variables are the coordinates of particles distributed in $k$
layers (or spins) not interacting with each other.

\smallskip

We generalize the construction to the case of a cylinder. The T.L.
representations acting on link patterns depend on a second
parameter $n'$ equal to the weight assigned to the loops which
wind around the cylinder.
We study the simplest case where $n'=2\cos{(\theta/ 2)}$, and in
the case where $n=n'=1$, we recover the stationary state of a
$O(1)$ model considered by Mitra and Nienhuis
\cite{nienhuis,nienhuis2}.

\bigskip

The specialization of these polynomials when all their variables
are set equal to one turns out to be polynomials in $n$ in the
disc case and $n'$ in the cylinder case, with integer
coefficients. We observe on examples and conjecture in general
that these integers are {\it positive}. When $n=n'=1$, the disc
and the cylinder polynomials become respectively the integers
previously conjectured to count alternating sign matrices and
half-turn alternating sign matrices in different topological
sectors \cite{razumov2}\cite{razumov3}.

In the simple case where the Hecke algebra representation is
generated from the action of difference operators on a product of
q-Vandermonde determinants, we have verified that the evaluation
at $z_i=1$ of the K-L basis polynomials are positive.

\bigskip
Although not directly related to these polynomials, we derive the
$c<1$ unitary modular invariant partition functions of conformal
field theories by decomposing certain representations of the
affine Temperley and Lieb algebra acting on Dynkin diagrams into
irreducible representations. As a byproduct, we define an action
of the modular group on the irreducible representations of the
affine T.L. (and more generally Hecke) algebra which relates them
to a tensor product of Virasoro representations.

\bigskip

The paper is organized as follows.

The first part is introductory and serves as a motivation, we
describe the representations of the affine T.L. algebra acting on
a system of lines projected onto the disk. We study in more detail
the case of the punctured disc.

The second and third part are the core of the paper and can to a
large extend be read independently. In the second part, we obtain
representations dual to those of the first part in terms of
polynomials obeying the wheel conditions. We give examples where
these polynomials are deformed Q.H.E. wave functions. In the third
part, we state the positivity conjectures when the variables of
the polynomials are set equal to one.

In the fourth part, we consider the case $n=1$, and we show that
the polynomials are the components of a $O(1)$ loop model ground
state. In the fifth part, we introduce the representations of the
affine T.L. algebra on SOS paths which diagonalize the affine
generators. In the sixth part, we consider more specifically the
unitary representation when the parameter of the algebra is a root
of unity. We decompose certain representations acting on Dynkin
diagram paths into irreducible representations.

\section{Representation of the affine T.L. algebra on patterns\label{T.L. algebra}}


The basic definitions of an affine Hecke algebra (A.H.A.)
$\mathcal{A}_N(t),$ and its T.L. restriction $\mathcal{A}^T_N(t),$
where $t$ is a parameter, are given in the appendix
\ref{section.hecke-algebra}. In this section, we define tangles
and patterns which are the natural objects to represent the action
of the generators of $\mathcal{A}^T_N$. The tangles carry the
three dimensional topological information. They can be decomposed
into a linear combination of planar tangles called patterns. The
patterns provide a basis of a representation of the affine T.L
algebra which we describe in this section.

\subsubsection{Tangles\label{section.string-pattern}}

A tangle is made by a set of open strings embedded into a three
dimensional ball such that their extremities are on the boundary.
We also consider the case where the ball is pierced by a flux
running through it. A string can eventually connect the center of
the ball or the flux to the boundary. The strings cannot cross
each other, they cannot cross the flux either. Two tangles are
considered equivalent if there is an ambient isotopy of one tangle
to the other keeping the boundary of the 3-ball (and the flux)
fixed.

\begin{figure}

\begin{center}

\includegraphics[width=6cm]{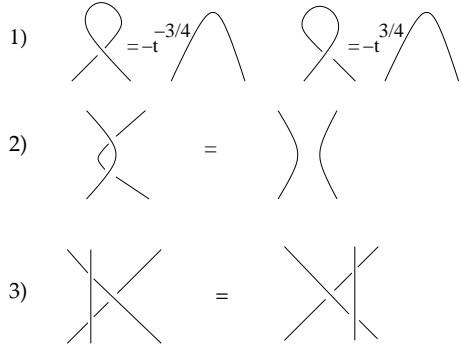}

\caption[99]{Reidemeister moves corresponding to the equations
(\ref{miscelanous}),(\ref{braid0}).}

\label{fig:reimaster1}

\end{center}

\end{figure}

The extremities of the strings on the boundary are arranged into
$N$  marked points $1,2,\dots ,N$ ordered anticlockwise around the
great circle. They are represented by their projection onto the
flat disc bounded by the great circle. By noting the under and
over crossings, one defines a tangle diagram. To respect isotopy
invariance, tangle diagrams are also identified through
equivalence relations known as Reidemeister moves. These moves are
represented on figure (\ref{fig:reimaster1}) (Strictly speaking,
the first Reidemeister move does not involve the factor
$-t^{-{3\over 4}}$).

In the punctured case we project the flux onto the origin and
therefore, the projected strings are not allowed to cross the
origin.

\subsubsection{Linear operators\label{section.linear-operators}}

Linear operators can be represented by an annulus with $N$
cyclically ordered marked points $(\bar 1,\dots ,\bar N)$ on its
inner boundary and $M$ cyclically ordered marked points $(1,\dots
,M)$ on its outer boundary connected by strings (a string can
connect two points of the same boundary). The action of the
annulus on the disk with $N$ marked points is null if $M\ne N$. It
is obtained by gluing its inner boundary to the boundary of the
disk so as to identify the marked points with those of the disk:
$(1,\dots ,N)=(\bar 1,\dots ,\bar N)$, and by joining the strings
of the disk with the strings of the annulus ending at the same
point. For example, the identity is represented by disjoined open
strings connecting $i$ to $\bar i$.

The braid group generators:
\begin{eqnarray}
 g_{i}&=&t^{-{1\over 4}}T_i=t^{{1\over 4}}+t^{-{1\over 4}}e_i,\cr
 g_{i}^{-1}&=&t^{{1\over 4}}T_i=t^{-{1\over 4}}+t^{{1\over 4}}e_i,
 \label{generators-braid}
 \end{eqnarray}
are represented by an annulus with $N-2$ disjoined strings
connecting $\bar l$ to $l$ for $l\ne k,k+1$, a string connecting
$\overline{k+1}$ to $k$, and a string connecting $\bar {k}$ to
$k+1$ passing over it (see figure \ref{fig:jones}).

The two first Reidemeister moves figure (\ref{fig:reimaster1})
result from the relations:
\begin{eqnarray}
 e_ig_i&=&-t^{-{3\over 4}}e_i\cr
 g_{i}g_{i\pm 1}e_i&=&e_{i\pm1}e_i\ \
 ,\label{miscelanous}
\end{eqnarray}
and the third move is the braid relation:
\begin{eqnarray}
  g_{i}g_{i+1}g_{i}&=& g_{i+1}g_{i}g_{i+1}\cr
  g_{i}g_{j}&=&g_{j}g_i\ \
 {\rm if} \ |i-j|>1.
 \label{braid0}
 \end{eqnarray}

\bigskip
A map from operators acting on the disc with $N$ marked points  to
operators acting on the disc with $N+1$ marked points consists in
adding an additional string to the pattern connecting
$\overline{N+1}$ to $N+1$ without adding any crossing.

Conversely, a partial trace is defined by joining together the
extremities $N$ and $\bar N$ without adding any crossings. It maps
operators acting on the disc with $N$ marked points to operators
acting on the disc with $N-1$ marked points.

\bigskip

\subsubsection{Patterns\label{section.link-pattern}}

\begin{figure}

\begin{center}

\includegraphics[width=8cm]{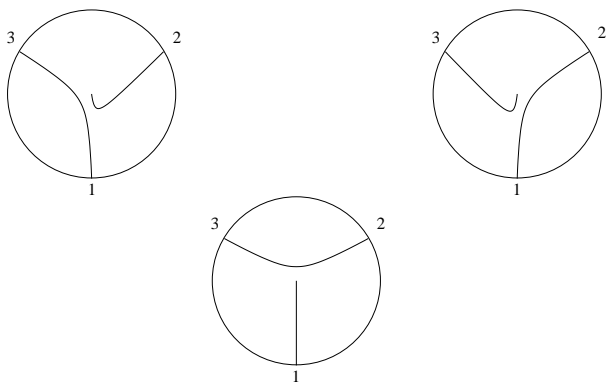}

\caption[99]{The patterns $\beta\alpha\alpha$,
$\alpha\alpha\beta$, $\alpha\beta\alpha$.}

\label{fig:string.1}

\end{center}

\end{figure}

A pattern is tangle with a flat projection on the disc. It can be
represented by disjoint lines connecting pairwise the boundary
points of the disk. Also, a line can start vertically down from
the inside to reach the boundary without winding (figure
\ref{fig:string.1}).

We can encode a pattern $\pi$ by a string of letters $\alpha$ or
$\beta$ \cite{lasc1}\cite{nienhuis}. We put $\alpha$ for $i$ if
the marked point is connected to the inside. In the disc case, we
put $\alpha$ for $i$ and $\beta$ for $j$ when $i<j$ are connected
by a line. Therefore, when we cut the string into two pieces, the
left piece must contain at least as many $\alpha$'s as $\beta$'s.
In the the punctured disc case, the lines are oriented so that the
domain bounded by it and the great circle surrounds the puncture
anticlockwise and we put $\alpha$ for the beginning and $\beta$
for the end of the line.

Given a string of $\alpha$'s and $\beta$'s, by successively
erasing factors $\alpha\beta$, one can recover the position of the
isolated $\alpha$'s.

It will often be convenient to view a pattern as an infinite
periodic string with the identification $\pi_{i+N}=\pi_i$.

We denote $\mathcal{H}_N$, the vector space made by linear
combinations of these patterns.

\begin{figure}

\begin{center}

\includegraphics[width=6cm]{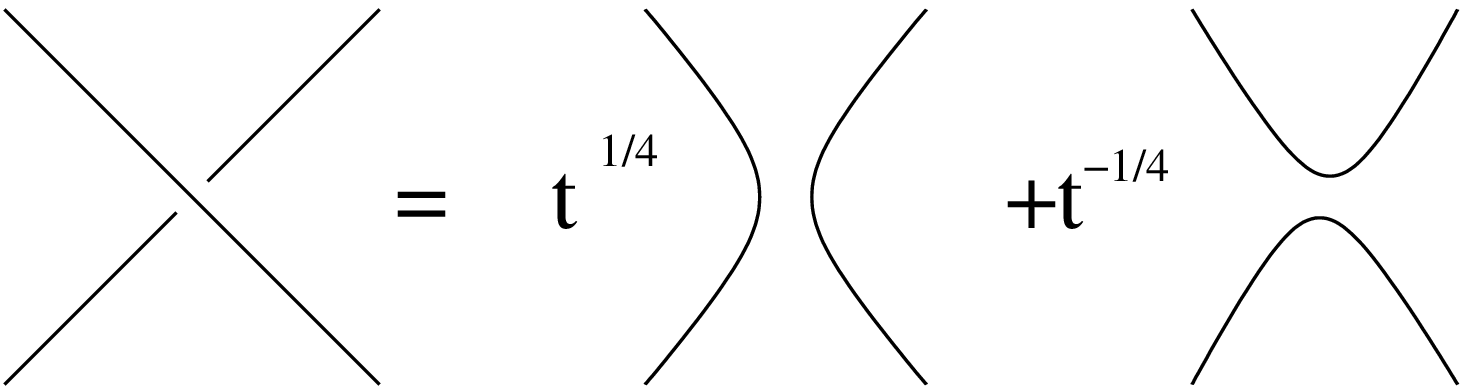}

\caption[99]{Skein relation corresponding to equation
(\ref{generators-braid}). The figure obtained by rotating each
piece by $90$ degree is also valid. }

\label{fig:jones}

\end{center}

\end{figure}

Tangles can be projected onto patterns as follows: Using the skein
relation of figure (\ref{fig:jones}), one can represent a tangle
with at least one crossing as a linear combination of two tangles
with one less crossing. A loop not surrounding the origin and not
crossing any other loop is removed by multiplying the weight of a
pattern by $\tau=-(t^{1\over 2}+t^{-{1\over 2}})$.

In the punctured disc case, if $N$ is even a loop surrounding the
origin and not crossing any other loop is removed by multiplying
it by $u+u^{-1}$, where $u$ is a new parameter. If $N$ is odd, we
require that if one rotates by an angle $ 2\pi$ around the origin
a line which starts from it, the weight is multiplied by $u$.
These transformations preserve the equivalence under Reidemeister
move, and using them, a tangle can be projected onto a linear
combinations of patterns.

We define linear operators acting in $\mathcal{A}^T_N$ as for
tangles. In particular, the T.L. generators $e_k$ are represented
by an annulus with $N-2$ disjoined lines connecting $\bar l$ to
$l$ for $l\ne k,k+1$, a line connecting $\bar k$ to
$\overline{k+1}$, and a line connecting ${k}$ to $k+1$. Using the
above rules, it is straightforward to verify the T.L. relations
with diagrams:
\begin{eqnarray}
 e_i^2&=&\tau e_i,\cr
 e_ie_j&=&e_je_i\ {\rm if}\ |i-j| \ge 2,\cr
 e_{i}e_{i\pm 1}e_{i}&=&e_{i}.
 \label{T.L.0}
\end{eqnarray}

\subsubsection{Temperley Lieb representations\label{Temperley Lieb
representations}}

We give the explicit expression of the T.L. matrices $e_i$ in the
link pattern basis.

\begin{figure}

\begin{center}

\includegraphics[width=6cm]{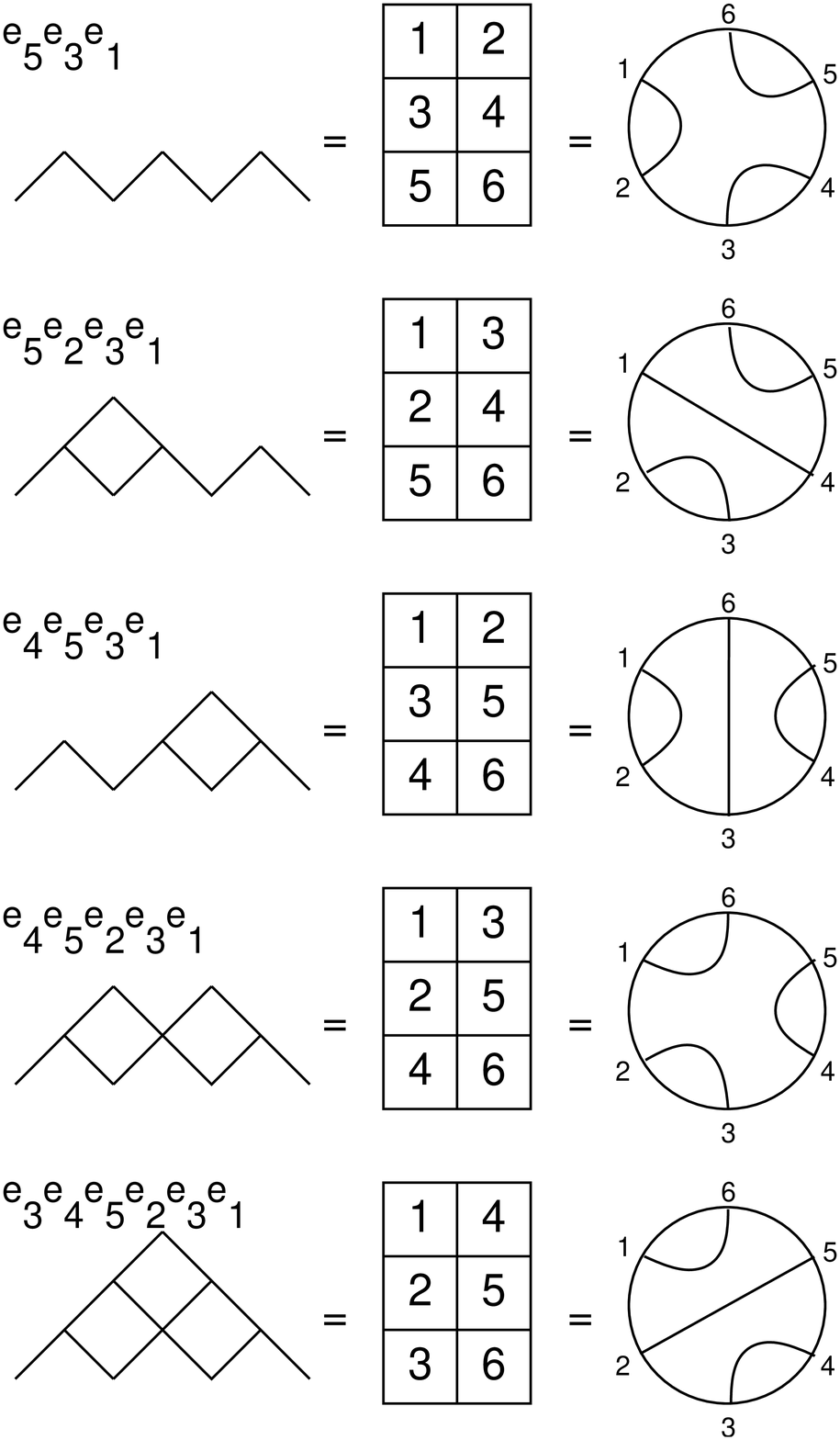}

\caption[99]{Several representations of the Hilbert space
$\mathcal{H}_6$ describing six particles on the disc, in terms of:

a) T.L. words or paths.

b) Young Tableaux where the two columns give the positions of the
$\alpha$'s and $\beta$'s of the string.

c) Patterns.

}

\label{fig:representation}

\end{center}

\end{figure}

We view a pattern as a string of $N$ letters of $\alpha$ and
$\beta$. If the difference  between the number of $\alpha$'s and
$\beta$'s is larger than or equal to two, it is not conserved
under the action of the T.L. algebra. Here, we restrict to the
case where this difference is equal to zero or one (see figure
\ref{fig:representation} for different representations of the
patterns).

Let us first consider the disc case. Following Lascoux and
Sch\"utzenberger \cite{lasc1}, we identify the link patterns basis
with a Kazhdan-Lusztig (K-L) basis (see section \ref{section-KL}).

\begin{figure}

\begin{center}

\includegraphics[width=6cm]{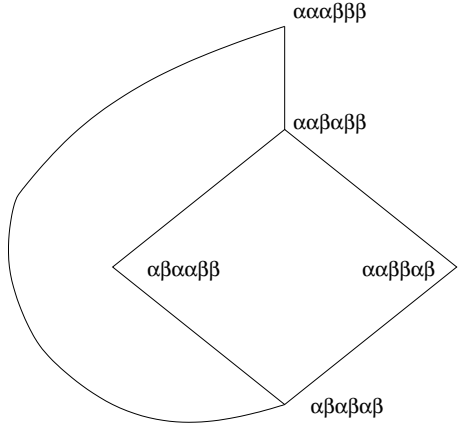}

\caption[99]{Kazhdan-Lusztig graph for 6 particles on the disc
($\mathcal{H}_6$).}

\label{fig:KL}

\end{center}

\end{figure}

A pattern $\pi$ carries the label $i$ if it is an eigenstate of
$e_i$, in other words if  $\pi_i\pi_{i+1}=\alpha\beta$.  Given a
pair of patterns $(\pi,\pi')$, we associate to it a reduced pair
$(\pi^r,\pi'^r)$, by successively erasing the factors
$\alpha\beta$ located at the same place of the two strings. Two
patterns are said to be matched if their reduced expression differ
only by a change of a factor $\alpha\beta\to\beta\alpha$. It can
be encoded into a K-L graph \cite{KL}, having the patterns as
vertices and an edge between two vertices when they are matched
(see figure \ref{fig:KL}). We define $\mu(\pi,\pi')=1$ or $0$
indicating if $\pi$ and $\pi'$ are matched or not.

The expression of the T.L. matrices $e_i$ is given by:
\begin{eqnarray}
 e_i\pi &=&\ \ \tau \pi,\ \ \ {\rm if}\ \pi\ {\rm has \ label\ } i,\cr
 &=&\sum_{\nu\ {\rm has \ label\ }i} \mu(\pi,\nu)\nu, \ {\rm if}\
 \pi\ {\rm has \ not\ label\ } i.
\label{lsei}
\end{eqnarray}

In the punctured disc case, this construction is modified as
follows. We view a pattern as an infinite periodic string of
letters $\alpha,\beta$. The reduction proceeds as in the disc case
where we take into account the periodicity of the patterns:
$\pi_{i+N}=\pi_i$. One still has the expression (\ref{lsei}) of
the T.L. matrices, where $\mu(\pi,\pi')=0$ if the patterns are not
matched and the value of $\mu(\pi,\pi')$ for matched pattern is
given by the following rules.

In the $N$ even case, $\mu(\pi,\pi')=1$, except if the two reduced
strings are $\alpha\beta$ and $\beta\alpha$ where
$\mu(\pi,\pi')=-(u+u^{-1})$.

In the $N$ odd case, $\mu(\pi,\pi')=1$, except if the two reduced
strings are $\pi^r=\beta\alpha\alpha$ and
$\pi'^r=\alpha\beta\alpha$ where $\mu(\pi,\pi')=u$,
$\pi^r=\beta\alpha\alpha$ and $\pi'^r=\alpha\alpha\beta$ where
$\mu(\pi,\pi')=u^{-1}$, and the permuted cases where
$\mu(\pi,\pi')=\mu(\pi',\pi)^{-1}$.

In the appendix \ref{appendix.one line}, we give the matrices
representing $\mathcal{A}^T_{2,3}$ which follow from the above
rules.

\subsubsection{Affine generators in the pattern representation
\label{section-Affine generators in the link pattern
representation}}

To obtain the affine algebra representation \cite{lehrer}, let us
define the cyclic operator $\sigma$ which rotates the ball by
$-2\pi/N$ around its axis. It acts on patterns by
 shifting the extremities of the strings by one unit clockwise: $i\to i-1$, so
that we have:
\begin{eqnarray}
\sigma g_i=g_{i-1}\sigma, \label{sigma-spin-def}
\end{eqnarray}
for $i\ge 2$, and one can define a new braid generator $g_N=\sigma
g_1 \sigma^{-1}$. In the punctured disc case, $\sigma$ acts by
shifting the indices $i\to i-1$.

Let us construct the affine generators $y_i$ which form a family
of commuting operators, and are therefore useful to characterize
the states of a representation. They also have a topological
interpretation.

\begin{figure}

\begin{center}

\includegraphics[width=6cm]{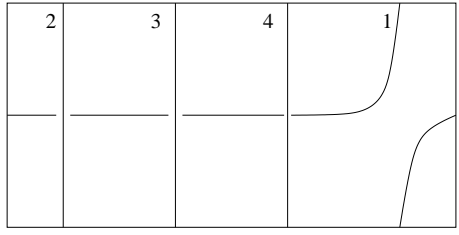}

\caption[99]{Affine generator $y_1$. }

\label{fig:affine}

\end{center}

\end{figure}

We define $y_1$ as:
\begin{eqnarray}
y_1=T_1T_2\dots T_{N-1}\sigma. \label{cycle1}\label{expression.y1}
\end{eqnarray}
Let us define $y'_1=t^{-{N-1\over 4}}y_1=\sigma g_2g_3\dots g_N'$.
It acts on tangles by letting the extremity of the string ending
at the marked point $1$ wind by an angle $-2\pi$ around the
boundary of the disk underneath the strings ending at position
$j\ne 1$ (see figure \ref{fig:affine}). Similarly, we define $y_i$
from the relations  defining the affine Hecke algebra
(\ref{affine-gener}) of appendix \ref{section.hecke-algebra}, and
$t^{2j-N-1\over 4}y_j$ lets the extremity of the string ending at
the marked point $j$ rotate by an angle $-2\pi$ around the
boundary of the disk above the strings ending at $j<i$ and
underneath the strings ending at $j>i$.

Let us consider the action of $y_1$ in the pattern representation.

If $N$ is even, the line connected to the marked point $1$ remains
underneath all the other lines and we can decouple it from them.
It is therefore sufficient to consider a pattern with only this
line. It is straightforward to verify that $y_1'=-t^{-{3\over 4}}$
if there is no puncture. In the punctured case, the basis has two
elements, $\beta\alpha$ and $\alpha\beta$, where the line
surrounds the puncture in two different ways and the matrix has
the expression (see appendix \ref{appendix.one line}):
\begin{eqnarray}
y'_1=\pmatrix{0&t^{{1\over 4}}\cr -t^{-{3\over
4}}&(u+u^{-1})t^{-{1\over 4}}}. \label{y1}
\end{eqnarray}
It therefore satisfies the quadratic relation:
\begin{eqnarray}
(y'_1-u^{-1}t^{-{1\over 4}})(y'_1-ut^{-{1\over 4}})=0.
\label{y1-cyclo}\label{y1-quadratique}
\end{eqnarray}
We define a second parameter $\tau'$ equal to the weight of a loop
surrounding the puncture. It is obtained by sandwiching $y'_1$
(\ref{y1}) between two $e_1$:
\begin{eqnarray}
 e_1y'_1e_1&=&-\tau't^{-{3\over 4}}e_1.
 \label{affine-gener-etrange}
\end{eqnarray}
Therefore, in the $N$ even case, we have:
\begin{eqnarray}
 \tau'=u+u^{-1}.
 \label{tau'-def-1}
\end{eqnarray}
In the case where $N$ is odd, we can proceed similarly by
considering a basis with three elements involving the patterns of
figure \ref{fig:string.1}. The matrix representing $y_1$ is given
in the appendix \ref{appendix.one line}. The relations
(\ref{y1-quadratique}) and (\ref{tau'-def-1}) are modified and
become:

\begin{eqnarray}
(y'_1-u^{-1}t^{-{1\over 2}})(y'_1-u)=0.
\label{y1-cyclo2}\label{y1-quadratique2}
\end{eqnarray}

\begin{eqnarray}
 \tau'=ut^{1\over 4}+u^{-1}t^{-{1\over 4}}.
 \label{tau'-def-2}
\end{eqnarray}

In both cases, $\tau'=x+x^{-1},$ where $x^2={y_+/ y_-}$ is the
ratio of the two eigenvalues of $y_1$.

\subsubsection{Involutions\label{section.Reflection-symmetry1 }}

One can define two natural involutions $\cal{F}$ and $ \cal{T}$ on
tangles.

The flip $\cal{F}$ rotates the ball by half a turn around a
horizontal axis passing through its center and relabels the points
$k \to N+1-k$. It induces the isomorphism taking: $g_k\to g_{N-k}$
and $\sigma \to \sigma^{-1}$.

The reflection $ \cal{T}$ reflects the ball through its equator.
It induces the antilinear involution taking: $g_k\to g^{-1}_{k}$,
$t\to t^{-1}$, and $\sigma$ to itself.




\subsubsection{Inclusion and Conditional expectation
value}\label{section-Conditional expectation value}

There is an imbedding of $\mathcal{H}_{N-2}$ into
$\mathcal{H}_{N}$ which takes $x$ to $e_1x$. On patterns, it
consists in shifting by two the labels of a pattern of
$\mathcal{H}_{N-2}$, and by adding a line connecting two new
points labelled $1$ and $2$ in such a way that $1,2,\dots,N$ are
cyclically ordered. Thus, it adds the two letters $\alpha\beta$ at
the beginning of the string representing the pattern.

Conversely, there is a projection $E$ called conditional
expectation value \cite{Jones} from $\mathcal{H}_{N}$ to
$\mathcal{H}_{N-2}$ defined by:
\begin{eqnarray}
e_1\pi=\tau E(\pi)e_1 .\label{project-2}
\end{eqnarray}
$E$ connects the lines ending at $1,2,$ so as to produce a pattern
starting with $\alpha\beta$, which is identified with a pattern in
$\mathcal{H}_{N-2}$ by the preceding imbedding. This projection is
hermitian for the scalar product previously defined and commutes
with the action of $\mathcal{A}^T_{N-2}$.

\section{Polynomial representations\label{dualpoly}}

We consider here polynomials in as many variables as there are
line extremities attached to the boundary of a pattern. One of our
aims is to identify representations of $\mathcal{A}^T_N$, dual to
those of the preceding sections, acting on these polynomials. This
is done through the introduction of a vector $\Psi$ whose
components are polynomials indexed by the patterns, and by
requiring that $\Psi$ transforms in the same way under the two
actions of the generators, on patterns or on polynomials. To
obtain these irreducible representations of $\mathcal{A}_N$, we
restrict the polynomials to obey some conditions called the wheel
conditions.

\subsection{The vector $\Psi$.\label{section.psi}}
The main problem of this section is to obtain a vector $\Psi$:
\begin{eqnarray}
\Psi=\sum_{\pi}\pi \psi_{\pi}(z_i),
\label{PDF}\label{psidefinition}
\end{eqnarray}
constructed in the following way. The vectors $\pi$ are basis
vectors of a representation on which $\mathcal{A}_N$ acts from the
left. The $\psi_{\pi}$ are polynomials on which it acts from the
right. We denote with a bar the right action of $\mathcal{A}_N$ on
polynomials to distinguish it from the left action. We want to
determine the $\psi_{\pi}$ in such a way that both actions give
the same result on the vector $\Psi$:
\begin{eqnarray}
\Psi\bar e_{i}&=& e_{i}\Psi \cr\Psi \bar\sigma&=&
\sigma\Psi,\label{duality}
\end{eqnarray}

Said differently, we identify the polynomials with the dual basis
$\psi_{\pi}(z_i)$ of the representation. These conditions are
equivalent to the conditions (\ref{intertwine-psi}) introduced
later.

Using a physical picture, $\psi_{\pi}(z_i)$ is the amplitude to
find the particles labelled by $i$ at positions $z_i$ in the
Resonance valance bond state (R.V.B.) $\pi$ where the lines
represent the spin singlets. The first condition (\ref{duality})
is the q-deformed conditions for $\Psi$ to be a bosonic wave
function.

For the disc and the punctured disc representations of section
\ref{section.link-pattern}, this problem is solved in section
\ref{subsection-T.L. representations}.




\subsection{The polynomials.\label{section.polynomes}}

We consider here Laurent polynomials in $N$ variables $z_i^{\pm
1}$ , $1\le i \le N$, which we identify with the generators $z_i$
of the double A.H.A. of appendix \ref{section.hecke-algebra}. We
exhibit a representation of the double A.H.A. on these polynomials
depending on two parameters $t$ and $q$.



We define permutation operators which permute the labels of the
variables $z_i$:
\begin{eqnarray}
 z_is_i&=&s_iz_{i+1},\cr z_{i+1}s_i&=&s_iz_i,\ {\rm and}\cr
 z_ls_{i}&=&s_iz_l\,\ {\rm if}\ l\ne i,i+1,
 \end{eqnarray}
Using the construction of appendix \ref{appendice-Permutation
operators},
one recovers the Lascoux and Sch\"utzenberger \cite{lasc2}
expression of Hecke generators $\bar e_i$ (for $1\le i\le N-1$)
acting from the right of an expression as follows:
\begin{eqnarray}
\bar e_i&=&(t^{1\over 2}z_{i+1}-t^{-{1\over 2}}z_{i})
(1-s_i)(z_{i}-z_{i+1})^{-1}\cr
 \bar e_i-\tau&=&(1-s_i)(z_i-z_{i+1})^{-1}{(t^{1\over 2}
 z_{i}-t^{-{1\over 2}} z_{i+1})},
 \label{polynomeTLei}
\end{eqnarray}
Therefore, $\bar e_i$ projects onto polynomials symmetrical under
the exchange of $z_i$ and $z_{i+1}$, and $\tau-\bar e_i$ onto
polynomials divisible by $(t^{1\over 2}z_{i}-t^{-{1\over
2}}z_{i+1})$. Notice that the generators $\bar e_i$ obey the
Hecke, not the T.L. relations.

\bigskip

It is convenient to define variables $z_i,\ i\in {\mathbb Z},$
which are cyclically identified as:
\begin{eqnarray}
z_{i+N}=q^{-1} z_i.\label{def-q}
\end{eqnarray}


We can represent the cyclic operator $\bar \sigma$ as:
\begin{eqnarray}
 { z_i\bar \sigma}&=&{ \bar \sigma z_{i+1},}
 \label{sigma-definition}
 \end{eqnarray}
and let it act as the identity when it is located left of an
expression:
\begin{eqnarray}
  1\bar\sigma &=&c_0.\ 
\label{sigmacycle1}
\end{eqnarray}
Thus, on a homogeneous polynomials of total degree $|\lambda|$,
$\bar\sigma^N=c_0^Nq^{-|\lambda|}$, where $c_0$ needs to be
adjusted  to satisfy the duality condition (\ref{duality}).

In the appendix (\ref{section.diag-yi})  we reproduce the results
of \cite{BGHP} showing that for the dominant ordering of the
monomials, the operators $\bar y_i$ are represented by triangular
matrices and we obtain their spectrum for $t,q$ generic. We
summarize the main results here:

\begin{itemize}
\item A partial ordering on monomials can be defined as follows
\cite{macdo2}.

Given a monomial $z^{\lambda}=z_1^{\lambda_1}\dots
z_N^{\lambda_N}$, we consider the partition $\lambda^+$ obtained
by reordering the $\lambda_i$ in decreasing order.

We say that $\lambda\ge\mu$ if the following conditions are
satisfied.

The partition $\lambda^+$ is larger than $\mu^+$ for the dominance
order defined as:
\begin{eqnarray}
\lambda^+_1+\dots\lambda^+_i\ge \mu^+_1+\dots\mu^+_i\ \ {\rm  for\
all}\ i\ge 1.
\end{eqnarray}

When $\lambda^+=\mu^+$,  $\mu$ can be obtained from $\lambda$
through a sequence of transformations
$(\lambda'_i,\lambda'_{i+1})\to (\lambda'_{i+1},\lambda'_{i})$
with $\lambda'_i>\lambda'_{i+1}$.

\item The polynomial representations of the A.H.A. depend on the
two parameters, $t$ and $q$, and on a partition $\lambda^+$.

The affine generators $y_i$ are lower triangular for the preceding
order.

An orthogonal basis is obtained by diagonalizing the $ y_i$
simultaneously and a basis state $F_{\pi}(z_i)$ is characterized
by its highest degree monomial:
\begin{eqnarray}
F_{\pi}(z_i) \propto z_{1}^{\lambda^+_{\pi_1}}\dots
z_{N}^{\lambda^+_{\pi_N}}+\dots ,
\end{eqnarray}
where $\pi$ is a permutation of $\lambda^+$:
$(\lambda_\pi)_i=\lambda^+_{\pi_i}$.

\item Up to an overall normalization factor, the eigenvalues of
the operators $ y_k$ on the polynomials $F_{\pi}$ are obtained by
permuting the eigenvalues $\hat y{_a}$ of $y_a$ on the highest
weight polynomial $F_{1}$ :
\begin{eqnarray}
 F_{\pi}y_k &=&F_{\pi}\hat y_{\pi_k} ,\ {\rm with:}\cr
 \hat y_a&=&q^{-\lambda^+_a}t^{(a-1)}.
\label{eigenvalues-y}
\end{eqnarray}
\end{itemize}

\subsection{Wheel condition \label{section.wheel}}

In this section, we introduce the vanishing conditions obeyed by
the polynomials which enable us to construct the vector $\Psi$.
These vanishing conditions, called the wheel conditions are
studied in \cite{kasatani}. When certain conditions are obeyed by
the parameters $t,q$, the space of polynomial obeying these wheel
conditions form a representation of the A.H.A.. We give their
definition and we explain why they are preserved under the action
of the A.H.A.. We motivate the vanishing conditions from the
Q.H.E. point of view by studying some examples in the next
section.

\subsubsection{Definition of the wheel conditions}

Fix two integers, $k$ and $r$, and two variables $t$ and $q$
related by:
\begin{eqnarray}
t^{k+1}q^{r-1}=1.
 \label{wheelcondition0}
\end{eqnarray}
If $m$ is the largest common divisor of $k$+1 and $r-1$, we take
$t^{k+1\over m}q^{r-1\over m}=\omega$, with $\omega$ a primitive
$m^{\rm th}$ root of unity.

We say that:

A Laurent polynomial $P(z_i)$, in $N$ variables $z^{\pm 1}_i$
satisfies the wheel condition $(k,r)$, if:
\begin{itemize}

\item For any subset of $k+1$ indices $\{i_a\},\ 1\le a\le k+1$.

\item For any set of $k$ integers $b_{aa+1}\in {\mathbb N},\ 1\le
a\le k$, such that:
\begin{eqnarray}
 &a)&\ \ b_{aa+1}=0 \ \Rightarrow \ i_{a+1}>i_{a},\cr
 &b)&\ \ \sum_1^{k} b_{aa+1}\le r-2.
 \label{wheelcondition1}
\end{eqnarray}

\item $P(z_1,\dots z_N)=0$ when we restrict the variables to
satisfy the wheel conditions:
\begin{eqnarray}
{z_{i_{a+1}}}=tq^{b_{a,a+1}}z_{i_{a}}.
 \label{wheelcondition2}
\end{eqnarray}
\end{itemize}

A set $\{i_a\}$,$\{b_{aa+1}\}$ satisfying the conditions
(\ref{wheelcondition1}) defines an admissible wheel, and the
vanishing condition specified by this wheel is called a wheel
condition.

We shall mostly be interested in the simplest cases  $r=2$ where
the rule simplifies drastically. One has $q=t^{-(k+1)}$ and given
$k+1$ ordered indices, $1\le i_1 \le i_2\dots \le i_{k+1}\le N$,
the polynomial must vanish when
$(z_{i_1},z_{i_2},\dots,z_{i_{k+1}})=(z,tz,t^2z,\dots,t^{k}z)$.

In the appendix \ref{section.proof stability} we show that the
wheel conditions are preserved under the action of the A.H.A.

\bigskip

\subsubsection{Admissibility conditions
\label{section.macdonald}}

Here, we recall the results of \cite{kasatani} about the
polynomial representations of the A.H.A. when the wheel conditions
are satisfied.

When the condition (\ref{wheelcondition0}), $t^{k+1}q^{r-1}=1,$ is
satisfied, and for $t$ generic, the representation admits an
irreducible subrepresentation on polynomials satisfying the wheel
condition (\ref{section.wheel}). The basis states are eigenstates
of the affine generators $y_i$ (\ref{expression.yi}) and are
proportional to the non-symmetrical Macdonald polynomials
\cite{BGHP}\cite{cherednick} specialized at $t^{k+1}q^{r-1}=1$.
They are characterized by their highest weight monomial now
subject to more restrictive admissibility conditions
\cite{kasatani}:
\begin{itemize}
\item A partition $\lambda^+=(\lambda^+_1,\dots,\lambda^+_N)$
defines an admissible state if it satisfies:
\begin{eqnarray}
\lambda^+_a-\lambda^+_{a+k}\ge r-1, \ \forall \ a\le N-k.
\label{condition-partition}
\end{eqnarray}
\item

A highest weight monomial is characterized by the shortest
admissible permutation $\pi$ such that one has:
$(\lambda_\pi)_i=\lambda^+_{\pi_i}$. A permutation $\pi$ defining
$\lambda_\pi$ is admissible if it satisfies the condition:
\begin{eqnarray}
\lambda^+_{a}-\lambda^+_{a+k}= r-1\Rightarrow a=\pi_i,\ a+k=\pi_j\
{\rm with}\ j>i. \label{condition-permutation-partition}
\end{eqnarray}
In other words, the weight
$\lambda_\pi=(\lambda^+_{\pi_1},\dots,\lambda^+_{\pi_N})$ is a
permutation of $\lambda^+$ such that $\lambda^+_{a}$ remains to
the left of $\lambda^+_{a+k}$ whenever
$\lambda^+_{a}-\lambda^+_{a+k}= r-1.$
\end{itemize}

In the the next section, we describe how the states of this
representation can be encoded into tableaux, and in section
\ref{section.paths}, we shall give a path description valid the
case of the two columns tableaux.

\subsubsection{Tableaux \label{appendix.Skew Young Tableau}}

The tableaux \cite{ram}\cite{suzuki} give a convenient way to
represent the states of the A.H.A. representations considered in
the last section.

An admissible permutation $\pi$ determining the polynomial
$F_{\pi}$ can be encoded by distributing the numbers $i,\ 1\le
i\le N$, into the boxes of a planar diagram. The boxes are
labelled from $1$ to $N$ and the polynomial $F_{\pi}$ is
represented by a tableau putting $i$ in the box $\pi_i$. We
identify the tableau representing the polynomial $F_\pi$ with the
permutation $\pi$.

The position of the box occupied by the number $i$ determines the
partial degree in the variable $z_i$ and the eigenvalue of the
affine generator $\bar y_i$ on $F_{\pi}$. Denote $(x_a,x'_a)$, the
cartesian coordinates of the box labelled $a$. Their sum labels
the eigenvalue $\hat y_a$ of $y_a$ (\ref{eigenvalues-y}), on the
highest-weight polynomial $F_{0}$:
\begin{eqnarray}
t^{x_a+x'_a}=\hat y_a,
 \label{boite=vp}
\end{eqnarray}
The product of the second coordinate with $r-1$, $(r-1)x'_a,$ is
the degree of the highest-weight polynomial $F_{0}$ in the
variable $z_a$. Thus:
\begin{eqnarray}
x_a&=&\lambda^+_a {k\over r-1}+a,\cr
 x'_a&=&{\lambda^+_a\over r-1}.
 \label{coordonnees-boite=vp}
\end{eqnarray}

As explained in the appendix \ref{appendice-Permutation
operators}, in order to obtain an irreducible representation, the
rule of construction is such that a vertical move of one unit
north or a horizontal move of one unit east has the effect to
multiply the eigenvalue $\hat y_a$ by a factor $t$.

The value of $x'_a$ modulo one, or equivalently the degree modulo
$r-1$, splits the boxes of a diagram into $r-1$ classes, and it is
convenient to split the diagram into $r-1$ disconnected
sub-diagrams.

The admissibility condition
(\ref{condition-permutation-partition}) can be rephrased into the
rule:

\begin{itemize}
\item The numbers are strictly increasing down each column and
across each row.
\end{itemize}
\bigskip

\begin{figure}

\begin{center}

\includegraphics[width=6cm]{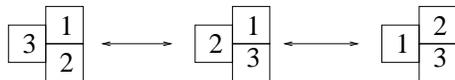}

\caption[99]{tableaux representing the states $(k,r)=(2,3)$
obeying the extended wheel condition $p=1$ for three particles.
The left box is disconnected from the right boxes.}

\label{fig:tableau}

\end{center}

\end{figure}

These are precisely the rules defining standard tableaux (see
figure \ref{fig:representation} and \ref{fig:tableau}).

Starting from any polynomial, one generates the others by acting
on it with intertwining operators $Y_i$ described in
\ref{appendice-Permutation operators}.
The tableaux are transformed into: $F_\pi\to F_\pi
Y_i=F_{\pi{s_i}}$ where $s_i$ exchanges the positions of $i$ and
$i+1$ in the tableau. $Y_i$ acts only if the two boxes are not
adjacent in the same row or column.

One can start from an arbitrary polynomial $F_{\pi}$ of the basis
to generate the other basis elements by acting with the $Y_i$ upon
it. In practise, the lowest degree polynomial has often a simple
factorized expression and is a more convenient generator. In the
next section, we exhibit some examples of these lowest degree
polynomials.

\bigskip

\subsection{Explicit solutions and Q.H.E. interpretation \label{Explicit solutions and Q.H.E.}}


We give some explicit solutions of the wheel conditions here.
Although we use the Q.H.E. terminology to motivate them from a
physical point of view, the polynomials of this section can easily
be obtained independently of any connection with the Q.H.E.

We consider particles moving in the plane in a strong magnetic
field projected to the lowest Landau level \cite{laughlin}. In a
specific gauge the orbital wave functions are given by:
\begin{eqnarray}
 \psi_{N}(z)= {z^k\over \sqrt{k!} } e^{- {z \bar z \over 4}},\label{orbitals}
\end{eqnarray}
where $z=x+iy$ is the coordinate of the particle, and the magnetic
length scale related to the strength of the magnetic field has
been set equal to one. These orbitals are concentrated on shells
of radius $\sqrt k$ occupying an area $2\pi $. Each orbital is
represented by a monomial $z^k$.

The quantum Hall effect \cite{read} ground state $\Psi$ is
obtained by combining these individual orbitals into a many-body
wave function. All the wave functions of system of $N$ particles
have a common factor $e^{-\sum_i {z_i {\bar z}_i \over 4 }}$ which
we omit. Thus, a monomial $z^{\lambda}=z_1^{\lambda_1}\dots
z_{N}^{\lambda_{N}}$ describes a configuration where the particle
$j$ occupies the orbital $\lambda_j$. The wave functions are
linear combinations of such monomials.

The physical properties are mainly characterized by the inverse
filling factor which is the area occupied by a particle measured
in units of $2\pi$. It can only be defined in the thermodynamical
limit, and is given by the limit when  $N\to \infty$ of the
maximum degree in each variable divided by the number of
variables.

The effect of the interactions is to impose some vanishing
conditions when the variables are in contact: $\Psi \sim
(z_i-z_j)^m$ with $m$ an integer when $z_i-z_j\to 0$. A ground
state wave function is a polynomial of the minimal degree obeying
the constraints. The difference between the polynomials considered
here and the Q.H.E. are the wheel conditions which reduce to the
Q.H.E. vanishing conditions when $t$ and $q$ tend to $1$ (in which
case, $k+1$ and $r-1$ must be prime numbers).
\bigskip

Through the examples considered here where the wave function has a
product structure, it turns out that $k$ has the interpretation of
a spin index and $r-1$ has the interpretation of an inverse
filling factor. The incompressibility condition translates into
obtaining the representations of section \ref{appendix.Skew Young
Tableau} having the most compact diagrams.


\begin{itemize}
{\item $k=1,$ solutions:}
\end{itemize}
Let us denote $\Delta_t(z_1,\dots,z_k)$ the Vandermonde product:
\begin{eqnarray}
\Delta_t(z_{1},\dots,z_{k})=\prod_{1 \le i<j \le {k}}(t^{1\over
2}z_i-t^{-{1\over 2}}z_j).
\end{eqnarray}
A solution obeying the $r$ wheel condition has the simple product
structure:

\begin{eqnarray}
 \psi_{k=1,r}(z_i)=\prod_{l=0}^{r-2}
\Delta_{tq^l}(z_{1},\dots,z_{N}).
 \label{Fomega.r}
\end{eqnarray}

It obviously satisfies the $k=1$, $r$ wheel conditions for wheels
with $i_1<i_2$, and if $i_2<i_1$, the wheel condition follows from
the fact that $(tq^{b_{12}})^{-1}=tq^{r-1-b_{12}}$.

This solution is the q-deformation of a Laughlin wave function
with an inverse filling factor $r-1$ \cite{read}.

Let us pursue the Q.H.E. analogy further. If we insert a magnetic
flux in the system at the origin, this has the effect to multiply
this wave function by a factor $\prod_i z_i$. Thus, the orbital
shells are expelled by one unit away from the origin and a region
of area $2\pi$ is left vacant, which is as if $1/(r-1)$ particle
had been removed from the origin. When the flux is inserted, the
eigenvalue of the operator $y_1$ gets multiplied by a factor
$q^{-1}$, which can be interpreted as the phase acquired by the
wave function when the particle winds around the flux.

\begin{itemize}
{\item $r=2,$ solutions with $k$ arbitrary:}
\end{itemize}
Let us show in the simplest case $r=2$ that $k$ has the
interpretation of a layer (spin) index. Consider $N$ particles
(variables) split into $k$ layers of $x_l,\ 1\le l\le k$ particles
each. We denote $z_{li},\ 1\le i\le N_l$ the coordinates of the
particles in the layer $l$ and we order the indices so that
${li}<{l'j}$ for $l<l'$. We say that two variables are in the same
layer if they share the same index $l$. The particles of the same
layer repel each other so that the wave function representing the
system vanishes when the variables $z_{lj}=tz_{li}$ for $i<j$, and
the particles belonging to different layers do not interact. A
ground state wave function representing this system obeys the
$(k,2)$ wheel condition because given $k+1$ variables, two of them
necessarily belong to the same layer.

A simple wave function obeying the vanishing conditions is thus
given by:
\begin{eqnarray}
\psi_{k,r=2}(z_i)=\prod_{l=1}^k \Delta_t (z_{l1},\dots,z_{lN_l}).
\label{Fomega.disc0.def}
\end{eqnarray}

Let us consider the incompressible limit. We look for polynomials
of the minimal degree obeying the wheel condition (the degree
measures the extension of the wave function). Thus, we split the
layers into two sets and set $k=k_1+k_2$. We fill each of the
$k_1$ first layers with $N-1$ particles and each of the last $k_2$
layers with $N$ particles. According to the rules of the preceding
section \ref{appendix.Skew Young Tableau}, the polynomials belong
to the irreducible representation characterized by the
skew-diagram, $k^N/1^{k_1},$ with $k_1$ columns of length $N-1$
and $k_2$ columns of length $N$. The polynomial
(\ref{Fomega.disc0.def}) coincides with the lowest weight of this
representation.

Notice that if we do not restrict to the minimal degree case and
allow the number of particles to differ by more than one in
between different layers, the representation is described by a
tableau with more than $k$ columns according to the rules of
section \ref{appendix.Skew Young Tableau}. Unless we specify it,
we shall not consider these cases here.

\begin{itemize}
{\item $k=2$, $r=3$ solution:}
\end{itemize}
The case $k=2$ and $r$, represents a system of two layers with an
inverse filling factor $r-1$, and for $r=3$, is the q-deformation
of the Haldane-Rezayi \cite{Haldane-Rezayi} wave function. We
repeat the Q.H.E. construction in the q-deformed case here. There
are $N=N_1+N_2$ particles or variables split into $2$ layers of
$N_1,N_2$ particles each. We label the variables by $li$ where
$l=1,2$ is the layer index and $i$ a particle index within each
layer. We order them so that $1i<2j$.

\smallskip

We split the wave function $\psi_{k=2,r=3}(z_i)$ into a product:
\begin{eqnarray}
\psi_{k=2,r=3}(z_{1i},z_{2j})=\phi^{}(z_{1i},z_{2j})\Delta_t(z_{11},\dots,z_{1N_1})
\Delta_t(z_{21},\dots,z_{2N_2}),
 \label{ansatz-psi0}
\end{eqnarray}
Due to the factors $\Delta_t$, the wheel condition is satisfied if
two indices $i_a,i_{a+1}$ involved in the wheel belong to the same
layer and $b_{aa+1}=0$. In particular, this covers all the cases
where the three variables belong to the same layer.

By inspection, the wheels left to be considered are those for
which:

$b_{aa+1}=0\Rightarrow $  $i_a$ is in the first layer and
$i_{a+1}$ is in the second layer.


Thus, if the two variables $1i$ and $2j$ belonging to different
layers participate to a wheel, the ratio $z_{2j}/z_{1i}$ is either
$t$ or $t^2q=(tq)^{-1}$. The last two are equal due to the
condition $t^3q^2=1$ (\ref{wheelcondition0}). Therefore, we must
impose $\phi$ to vanish in these cases.

For $\phi{},$ we consider an expression of the form:
\begin{eqnarray}
\phi(z_{1i},z_{2j})=\prod_{i=1}^{N_1}\prod_{j=1}^{N_2}
f(z_{1i},z_{2j})\prod_{k=1}^{{\rm inf}\{N_1,N_2\}}
f^{-1}(z_{1i_k},z_{2i_k}),
 \label{ansatz-psi1}
\end{eqnarray}
where $1i_k,2i_k$ is a maximal pairing between particles of the
the first layer with those of the second layer and $N_1-N_2$
particles of the first (second) layer are not paired if $N_1>N_2$
($N_2>N_1$). For $f$ we take:
\begin{eqnarray}
f(z,w)=(t  z-w)(z-t q w).
 \label{f-expression}
\end{eqnarray}

In all the wheel cases considered above, two of the three
particles participating to the wheel, say $1i$ and $2j$, belong to
different layers and are unpaired. Since the wave function $\phi,$
contains the factor $f(z_{1i},z_{2j})$ it vanishes for this wheel.
Therefore, the wheel condition is always satisfied.

The degree of $\phi$ can be reduced by antisymmetrizing it over
the variables of the same layer $z_{li}$ and by dividing the
result by $\prod_{l=1}^2\prod_{i<j}(z_{li}-z_{lj})$.

If we restrict to the minimal degree cases $N_1=N_2=N$ or
$N_1=N-1,N_2=N$, using the rules of the preceding section
\ref{appendix.Skew Young Tableau}, we see that the polynomial
obtained here is the lowest weight state of the representations
characterized by the (skew)-diagram, $2^N$ or $2^N/1$, with two
columns of length $N_1$ and $N_2$.

When $N_1=N_2,$ the wave function is proportional the Gaudin
Determinant \cite{gaudin}:
\begin{eqnarray}
 \psi_{k=2,r=3}^{N_1=N_2}={\rm Det}{\big (}{1\over (t^{1\over 2}  z_{1i}-t^{-{1\over 2}}z_{2j})
 (t^{1\over 4}z_{1i}-t^{-{1\over 4}}z_{2j})}{\big )}
 {{\Delta_t(z_{11},\dots,z_{1r})\Delta_t(z_{21},\dots,z_{2r})}\over \prod_{i<j} (z_{1i}-z_{1j})\prod_{i<j}
 (z_{2i}-z_{2j})}.
 \label{f-expression-gaudin}
\end{eqnarray}


\smallskip

For $r> 3$, we were not able to obtain a simple trial polynomial
obeying the $(2,r)$ wheel condition with the minimal degree. The
polynomials obeying these wheel conditions can nevertheless be
obtained by specializing the non-symmetric Macdonald polynomials
at $t^{k+1}q^{r-1}=1$.



\subsubsection {Inserting a flux\label{section flux}}

The polynomials with $r>2$ allow to describe tangles with a flux
inserted.

Let us fix an integer $p$ with $1\le p\le r-1$ and impose the
additional wheel condition:

\begin{itemize}
{\item The polynomials vanish at least as $\epsilon^p$ when two
arbitrary variables $z_i$ and $z_j$ approach zero as $\epsilon$.}
\end{itemize}

This constraint is preserved under the action of the A.H.A.. It
can be realized on the polynomial representations characterized by
a diagram with two disjoined columns vertically shifted by
$p/(r-1)$ with respect to one-another.

This results from the fact that given a polynomial in this
representation and a monomial in its expansion, the sum of any two
of its degrees is greater or equal to $p$. This is true for the
highest monomial of an eigenstate of the affine generators $\bar
y_i$, and thus for any monomial in its expansion by the
construction of the appendix \ref{section-polynomial
representation}.

The $p=1$ case can simply be obtained by multiplying the disc
polynomial $\psi_{k=2}^{N_1,N_2}$ by the product
$\prod_{i=1}^{N_l} z_{li}$ where $l\in \{1,2\}$ is the column with
the smallest length.

 \subsection {Link pattern basis\label{subsection-T.L. representations}}

Here, we identify the polynomial basis dual to the pattern basis
of section (\ref{section.link-pattern}). In the disc case, the
basis coincides with the K-L basis and in the punctured disc case,
the construction needs to be modified.
\bigskip

\subsubsection{Kazhdan-Lusztig basis\label{section-KL}}

We first give the general construction of the K-L basis \cite{KL}
which we then specialize to the case of the the disc patterns
\cite{lasc1}.

The generating polynomial is a product of q-Vandermonde
determinants as in (\ref{Fomega.disc0.def}):
\begin{eqnarray}
\psi_1 (z_i)=\prod_{l=1}^k \Delta_t (z_{l1},\dots,z_{lN_l}),
\label{Fomega.disc0.def2}
\end{eqnarray}
and the $N_l$ are sorted decreasingly. It is more convenient here
to index the variables from $1$ to $N=\sum_{l=1}^k N_l$ with
$li\to N_1+\dots N_{l-1}+i$.


We construct a representation of the Hecke Algebra called Specht
representation \cite{fulton} by letting the generators $e_i-\tau$
(\ref{polynomeTLei}) with $1\le i\le N-1$ act on this polynomial.
It has a K-L basis which we describe here.

The basis states are given in terms of standard tableaux
associated to a Young diagram as in section \ref{appendix.Skew
Young Tableau}. The lowest tableau, denoted $1,$ has its columns
filled with consecutive numbers and corresponds to the polynomial
$\psi_{1}$ (It corresponds to the lowest tableau of figure
\ref{fig:representation}) . We number the boxes of diagram as in
$1$. The standard tableau $\pi$ puts the number $i$ in the box
$\pi_i$. It is generated from the tableau $1$ by a succession of
elementary permutations $\pi\to \pi s_i$, where $s_i$ exchanges
the position of $i$ and $i+1$, and acts only if $i+1$ is in a
column to the right of $i$ and not in the same line as $i$. This
determines an order on tableaux, the dominance (or Bruhat) order:
$\pi'\ge\pi$ if it can be written in the preceding way as
$\pi'=\pi w$ with $w$ a word in the $s_i$.

To the tableau $\pi$, we associate an element $\bar T_\pi$ in the
Hecke algebra obtained by substituting the generator $\bar T_i$ to
the permutation $s_i$ in the above expression of $\pi$. We obtain
a polynomial basis of the Specht module, labelled by tableaux, by
letting $\bar T_{\pi}$ act on $\psi_{1}$:

\begin{eqnarray}
 F_\pi (z_i)=\psi_1\bar T_{\pi},
 \label{Fomega.disc0.def3}
\end{eqnarray}

We define an antilinear (but preserving the order) involution by
${\cal T}' ( \bar T_i)=\bar T_i^{-1}$, $t^{1\over 2}\to
t^{-{1\over 2}}$ and ${\cal T}' ( \psi_{1})=\psi_{1}.$ The K-L
basis states $\psi_{\pi}$ are obtained from the basis states
$F_{\pi}$ by a triangular transformation and are defined by the
relations:
\begin{eqnarray}
 {\cal T}' (\psi_{\pi})&=&\psi_{\pi}\cr
\psi_{\pi}-F_{\pi}&\in& \oplus_{\pi'<\pi} t^{-{1\over 2}} {\mathbb
Z}[t^{-{1\over 2}}]F_{\pi'}.
 \label{KLrelations}
\end{eqnarray}

Let us make a few observation about the K-L basis.

\smallskip

The Specht representation coincides with the AHA representation of
section \ref{appendix.Skew Young Tableau} only in the minimal
degree case where the length of the columns of the Young diagram
differ by at most one. In this case the cyclic permutation
$\bar\sigma$ (\ref{sigma-definition}) acts within the
representation.
\smallskip

If we denote by $J$ the subset of $\{1,\dots N\}$ where we omit
the numbers $\{N_1,N_1+N_2,\dots,N\}$, the polynomial
(\ref{Fomega.disc0.def2}) obeys:
\begin{eqnarray}
 \psi_{1}\bar T_i=t^{1\over 2}\psi_1,\ \forall i\in J.
 \label{KL1}
\end{eqnarray}

One can define a K-L basis for the induced module defined by the
relation (\ref{KL1}). The Specht representation is a
subrepresentation of this induced module.

One can define a KL-graph as in section \ref{section-KL} and the
expression (\ref{lsei}) of the matrices $e_i$ holds in the general
case. This gives a practical way to construct the K-L basis. The
tableau $\pi$ carries the label $i$ if $i$ is in a column to the
left of $i+1$. An incomplete K-L graph (The Young graph) is drawn
by connecting $\pi$ to $\pi'$ when $\pi$ and $\pi'$ differ by the
permutation of two consecutive numbers. The trial basis states are
obtained recursively by acting with $\bar e_i-\tau$ on
$\psi_{\pi}$ carrying the label $i$ and such that $i$ and $i+1$
are not in the same line:
\begin{eqnarray}
 \psi_{\pi
s_i}=\psi_{\pi}(\bar e_i-\tau)-\sum_{\nu<\pi\ {\rm without\ label\
}i} \mu(\pi,\nu)\psi_{\nu}.
\label{lsei-trial}
\end{eqnarray}
Using the first relation (\ref{duality}), one constructs the
matrices $e_i-\tau$ in the trial basis and one completes the
KL-graph using (\ref{lsei-trial}). Due to the factor $t^{1\over
2}$ in (\ref{KL1}), the basis may violate the second condition in
(\ref{KLrelations}) (equivalently, $e_i-\tau$ violates the
non-negative integrality condition (\ref{lsei})). This is cured by
completing the K-L graph with new links and by correcting the
basis states (\ref{lsei-trial}) accordingly.

\smallskip

The dual basis is also a K-L basis. It is generated by induction
from the highest tableau having its lines filled with consecutive
numbers. In the dual construction, $t^{-{1\over2}}$ must be
substituted to $t^{1\over 2}$ in the relations (\ref{KLrelations})
and (\ref{KL1}) and the conjugate diagram must be substituted to
the original diagram. So, the set $J$ is replaced by $J'=\{1,\dots
N\}\setminus \{N'_1,N'_1+N'_2,\dots,N\}$ where the $N'_i$ are the
lengths of the lines.

\bigskip\bigskip

Specializing to the two column diagrams, we identify the tableaux
with the link patterns of section \ref{section.link-pattern}. A
tableau is the link patterns where the numbers in the left column
encode the position of the $\alpha$'s, and the numbers in the
right column the positions of the $\beta$'s as in figure
\ref{fig:representation}. The lowest tableau is the pattern having
all the $\alpha$'s precede the $\beta$'s.

Two strings are connected by a link of the Young graph if they
differ by a simple transposition of two consecutive $\alpha$,
$\beta$, and new links must be added to obtain the KL-graph
according to the rules given in section \ref{Temperley Lieb
representations}. For example, on figure \ref{fig:KL}, the KL-link
connecting $\alpha\alpha\alpha\beta\beta\beta$ to
$\alpha\beta\alpha\beta\alpha\beta$ is not a link of the Young
graph .

By transposing the relations (\ref{lsei}) we obtain the action of
$\bar e_i- \tau$ on the basis states:
\begin{eqnarray}
 \psi_\pi (\bar e_i-\tau) &=&\ \ \ \ \ \ \ -\tau \psi_\pi\ \ \ \ {\rm if}\ \pi_i\pi_{i+1}\ne\alpha\beta,\cr
&=&\sum_{ \nu_i\nu_{i+1}\ne\alpha\beta}\mu(\nu,\pi)\psi_\nu, \
{\rm if}\ \pi_i\pi_{i+1}= \alpha\beta,
 \label{lseit}
\end{eqnarray}
which give a convenient way to generate the
basis.\footnote{Another way to proceed is to use the factorized
expression of the basis states in terms of a product of
Yang-Baxter operators acting on the lowest tableau
\cite{lasc-kiril}.}

In the case where the number of $\beta$ is equal to the number of
$\alpha$, the K-L involution ${\cal T}'$ can be identified with
the reflection ${\cal T}$ of the section
\ref{section.Reflection-symmetry1 }.

\subsubsection{Punctured disc basis\label{punctured disc}}

Our aim here is to construct the basis of polynomials dual to the
punctured disc patterns of \ref{section.link-pattern} using the
flux representation considered in \ref{section flux}.

The representation cannot be generated using the Hecke algebra
only and one must add the generator $\sigma$. We must enlarge the
involution to include $\sigma$ and we require that ${\cal
T}'(\sigma)=\sigma$.

As a generator of the representation, we take the lowest patterns
$\omega=\beta\dots \alpha$ with the $\beta$'s preceding the
$\alpha$'s. This pattern is not in the image of any $e_i$, and
therefore, the polynomial $\psi_{\omega}$ must be annihilated by
the $\bar e_i$ acting from the right. The minimal degree candidate
is given by:
\begin{eqnarray}
 \psi_{\omega}(z_i)=\Delta_t (z_1,\dots,z_N).
 \label{Fomega.cylindre.def}
\end{eqnarray}

This polynomial has a highest degree monomial given by
$z_1^{N-1}z_2^{N-2}\dots z_N^0$. Thus the partition of this
representation is given by:
\begin{eqnarray}
 \lambda^+=N-1,N-2,\dots,0.
 \label{Fomega.cylindre-partition}
\end{eqnarray}
$\psi_{\omega}$ obeys the wheel condition $k=2,\ r=3,$ which
resumes to require that for triplet $i<j<k$ the polynomial
vanishes when $z_j=tz_i$ or $z_k=tz_i$ or $z_k=tz_j$. It also
vanishes when two variables are set equal to zero. It therefore
belongs to flux representation $p=1,r=3$ of section \ref{section
flux}.

The construction of the basis proceeds as in the disc case where
we use the generators
$\psi_{\omega\sigma^l}=\psi_\omega\bar\sigma^l$ to obtain the
other basis states using (\ref{lseit}) in each sector determined
by $l$.


\smallskip

To characterize the pattern representation, we need to determine
the parameter $\tau'$ (\ref{affine-gener-etrange}). The ratio
$y_+/y_-$ of the two possible eigenvalues of $y_1$, obtained from
(\ref{ur-us}) and from (\ref{eigenvalues-y}) must be equal. It
gives, $q^{-1}t^{-1}=u^2$ if $N$ is even, and
$q^{-1}t^{-1}=u^2t^{1\over 2}$ if $N$ is odd, where $q=t^{-{3\over
2}}$ by the condition (\ref{wheelcondition0}). Thus,
\begin{eqnarray}
 u&=&t^{1\over 2}\ {\rm if}\ n\  {\rm is \ even},\cr
 u&=&1\ {\rm if}\ n\ {\rm is \ odd}.
\label{expr-u}
\end{eqnarray}
Therefore, in this representation, and if we parameterize
$\tau=-2\cos{(\theta)}$,  we obtain from (\ref{tau'-def-1}) and
(\ref{tau'-def-2}):
\begin{eqnarray}
 \tau'=2\cos{({\theta\over 2})}.
\end{eqnarray}
We notice that in all cases, $\sigma^N=1$. So, in order to satisfy
the second duality relation (\ref{duality}), the normalization
factor of $\bar \sigma$ must be taken equal to:
$c_0=q^{|\lambda|\over n}$ in (\ref{sigmacycle1}).

\subsubsection{Conditional expectation
value and  Involutions\label{section-Conditional expectation
value2}\label{section.Reflection-symmetry2}}

We give two properties of these polynomials parallel to the link
patterns properties discussed in sections
\ref{section.Reflection-symmetry1 } and \ref{section-Conditional
expectation value}:

\smallskip
One can realize the involutions dual to those of section
(\ref{section.Reflection-symmetry1 }) through the transformations:

\begin{eqnarray}
\bar {\cal  F}(\Psi)(z_1,\dots,z_N)&=&\Psi({
z_N^{-1}},\dots,z_1^{-1}).\cr \bar {\cal
T}(\Psi_t)(z_1,\dots,z_N)&=&\Psi_{t^{-1}}( z_1^{-1},\dots,
z_N^{-1}).
 \label{reflection-isomorpism2}
\end{eqnarray}
\bigskip

The projection $E'$ dual to the inclusion defined in
(\ref{project-2}) is defined in the appendix
(\ref{section-Conditional expectation value in the T.L}) by
specializing the two first variables to take the value $1$ and
$t^2$. It sends polynomials in $N$ variables obeying the vanishing
condition into polynomials in $N-2$ variables obeying the same
property. By combining it with the cyclic transformation
(\ref{sigma-definition}), it gives a way to decompose a polynomial
satisfying the wheel condition in the link pattern basis by
specializing its variables to be powers of $t$.

\section{Positivity conjectures \label{section.conjectures}}

The positivity conjectures are motivated by the R-S conjecture
\cite{razumov2} which states that at $\tau=1$, the evaluations of
the polynomials considered in the preceding section at $z_i=1$
count certain classes of alternating sign matrices. We claim that
the evaluation of their deformations are positive polynomials in
the deformation parameter $\tau$ (are in $\mathbb{N}[\tau])$.

Similarly, we observe that the evaluation of the cylinder
polynomials are in $\mathbb{N}[\tau']$, where where we set
$\tau=-2\cos{(\theta)}$ and $\tau'=2\cos{(\theta/2)}$. When
$\tau=\tau'=1$, these polynomials count certain classes of
half-turn symmetric alternating sign matrices \cite{razumov3}.

\smallskip

A first observation concerns the K-L basis constructed from the
product of Vandermonde determinants (\ref{Fomega.disc0.def2}). We
define the evaluation of a polynomial to be:
\begin{eqnarray}
 \bar\psi_{\pi}&= & N^{-1}\psi_{\pi}(1,\dots,1),
 \label{evaluation}
\end{eqnarray}
where the normalization factor $N=\bar \psi_1(1,\dots,1)$ in
(\ref{Fomega.disc0.def2}).

Let us show that the evaluation of K-L basis polynomials
$\bar\psi_{\pi}$ are in $\mathbb{Z}[t^{\pm{1\over 2}}]$. The
$\psi_{\pi}$ are polynomials in $z_i$ with coefficients in
$\mathbb{Z}[t^{\pm{1\over 2}}]$. The property is true for
$\psi_1$, and preserved under the action of $\bar e_i-\tau$
(\ref{polynomeTLei}). Thus,
$\bar\psi_{\pi}\in\mathbb{Z}[t^{\pm{1\over 2}}]$ if
$\psi(1,\dots,1)$ is divisible by $(t-1)^{\sum_l{ N_l(N_l-1)\over
2}}$. This results from the fact that when $t-1$ and $z_i-1$ are
$O(\epsilon)$ $\forall i,$ $\psi_{\pi}$ is $O(\epsilon^{\sum_l{
N_l(N_l-1)\over 2}})$. Again, the property is satisfied by
$\psi_1$ and preserved by the action of $\bar e_i-\tau$ (and $\bar
\sigma$).

We have verified that they are in fact {\it positive} in
$-t^{\pm{1\over 2}}$.



\bigskip

Let us consider more specifically the link pattern polynomials.

The disc polynomials (without line connected to the center) are
obtained by considering K-L representations having a Young tableau
with two columns of equal length.

In this case, $\bar\psi_{\pi}$ is invariant under the
transformation $t^{1\over 2}\to t^{-{1\over 2}}$ and is therefore
in $\mathbb{Z}[\tau]$. To show this, let us consider the
reflection ${\cal T}$ which takes $z_i\to z_i^{-1}$ and $t^{1\over
2}\to t^{-{1\over 2}}$ and multiplies the result by a factor
$(z_1\dots z_N)^{({N\over 2}-1)}$. If a polynomial $\psi$
invariant under a ${\cal T}$ reflection, then $\bar\psi$ is
invariant under the transformation $t^{1\over 2}\to t^{-{1\over
2}}$. ${\cal T}$ leaves $\psi_1$ invariant and commutes with $\bar
e_i-\tau$ (and  $t^{-{3\over 2}({N\over 2}-1)}\bar \sigma$).
Therefore, the $\psi_{\pi}$ are left invariant under ${\cal T}$
and the property $\bar \psi_\pi\in\mathbb{Z}[\tau]$ follows.

We conjecture that the evaluation of the disc polynomials are
positive in $\tau$.

In table \ref{tab1}, we evaluate explicitly the polynomials $\bar
\psi_\pi$ up to $N=8$, and we observe that $\bar
\psi_\pi\in\mathbb{N}[\tau]$.

\begin{table}[h]
\begin{tabular}{|l |l|}

\hline 12 & $1$\\
\hline 13 & $\tau$ \\ \hline
\end{tabular}
\begin{tabular}{|l |l|}

\hline 123 & $1$\\
\hline 124 & $2\tau$ \\
\hline 134 & $\tau^2$\\
\hline 125& $\tau^2$\\
\hline 135 & $\tau(\tau^2+1)$\\ \hline
\end{tabular}
\begin{tabular}{|l |l|}

\hline 1234 & $1$\\
\hline 1235 & $3\tau$ \\
\hline 1245  & $3\tau^2$\\
\hline 1236  & $3\tau^2$\\
\hline 1345 & $\tau^3$\\
\hline 1237  & $\tau^3$\\
\hline 1246 & $\tau(5\tau^2+2)$\\
\hline 1256 & $\tau^4$\\
\hline 1346  & $\tau^2(2\tau^2+1)$\\
\hline 1247  & $\tau^2(2\tau^2+1)$\\
\hline 1356  & $\tau^3(\tau^2+2)$\\
\hline 1257  & $\tau^3(\tau^2+2)$\\
\hline 1347 & $\tau(\tau^4+\tau^2+1)$\\
\hline 1357 & $\tau^2(\tau^4+3\tau^2+3)$\\ \hline
\end{tabular}
\caption{ Evaluation of the disc polynomials  for $k=2,\ r=2$ and
$N=4,6,8$. We index the polynomials $\bar\psi_\pi$ by the position
of the $\alpha$'s in the notation of \ref{section.link-pattern}.
The evaluation is left unchanged under the flip isomorphism
defined in \ref{section.Reflection-symmetry1 } and
\ref{section.Reflection-symmetry2}: $\bar \psi_\pi=\bar
\psi_{{\cal  F}(\pi)}$. } \label{tab1}
\end{table}

In table \ref{tab2}, we evaluate the polynomials $\bar \psi_\pi$
with $k=2,r=3,$ and $N=2,4$, obtained from the polynomial
$\psi^{(n)}_{2,3}$ (\ref{ansatz-psi0}). The normalization factor
in (\ref{evaluation}) is taken to be $ (-)^{N\over 2}(t^{1\over
4}-t^{-{1\over 4}})^{N({N\over 2}-1)}(t^{1\over 4}+t^{-{1\over
4}})^{{N\over 2}({N\over 2}-1)}$. This time, we observe that the
evaluation polynomials are positive in the variable
$\tau'^2=2-\tau$.

\begin{table}[h]
\begin{tabular}{|l |l|}

\hline 12 & $1+\tau'^2$\\
\hline 13 & $2$ \\ \hline
\end{tabular}
\begin{tabular}{|l |l|}

\hline 123 & $\tau'^6+2\tau'^4+3\tau'^2+1$\\
\hline 124 & $2\tau'^4+8\tau'^2+4$ \\
\hline 134 & $3\tau'^2+4$\\
\hline 125 & $3\tau'^2+4$\\
\hline 135 & $\tau'^4+3\tau'^2+10$\\ \hline
\end{tabular}
\caption{Evaluation of the disc polynomials  for $k=2,\ r=3$ and
$N=4,6$. } \label{tab2}
\end{table}


In the punctured disc case, we have considered the simplest cases
where the representation is generated from the Vandermonde
determinant (\ref{Fomega.cylindre.def}) (which corresponds to the
case $k=2,r=3,p=1,$ in the notations of section
\ref{subsection-T.L. representations}).  The normalization factor
in (\ref{evaluation}) is taken to be $N=(-)^{N\over 2}(t^{1\over
4}+t^{-{1\over 4}})^{^{{N\over 2}({N\over 2}-1)}}(t^{1\over
4}-t^{-{1\over 4}})^{N(N-1)\over 2}$ if $N$ is even, and
$N=(t^{1\over 4}+t^{-{1\over 4}})^{{(N-1)^2\over 4}}(t^{1\over
4}-t^{-{1\over 4}})^{N(N-1)\over 2}$ if $N$ is odd. One can show
the same integrability conditions as in the disc case with
$t^{1\over 2}\to t^{1\over 4}$. The explicit evaluations of table
\ref{tab3} lead us to conjecture that the evaluations are positive
in $\tau'$.

\begin{table}[h]
\begin{tabular}{|l |l|}
\hline $1$ & $1$ \\
\hline $2$& $\tau'$\\\hline
\end{tabular}
\begin{tabular}{|l |l|}
\hline $12$ & $1$ \\
\hline $13$ & $1$ \\
\hline $23$& $\tau'^2$\\\hline
\end{tabular}
\begin{tabular}{|l |l|}

\hline  12& $1$\\
\hline  13& $\tau'^2+2$ \\
\hline  23& $\tau'$\\
\hline  14& $\tau'$\\
\hline 24& $\tau'(\tau'^2+2)$\\
\hline 34& $\tau'^4$\\\hline

\end{tabular}
\begin{tabular}{|l |l|}

\hline 125  & $1$\\
\hline 123  & $1$\\
\hline 145 & $\tau'^2$ \\
\hline 234 & $\tau'^2$ \\
\hline 135 & $\tau'^2+3$\\
\hline 124 &$\tau'^2+3$\\
\hline 245& $\tau'^2(\tau'^2+3)$\\
\hline 235& $\tau'^2(\tau'^2+3)$\\
\hline 134& $\tau'^4+\tau'^2+2$\\
\hline  345& $\tau'^6$\\\hline

\end{tabular}

\begin{tabular}{|l |l|}

\hline 123  & $1$\\
\hline 124  & $\tau'^2+4$\\
\hline 125  & $(\tau'^2+2)^2$\\
\hline 134 & $(\tau'^2+2)^2$\\
\hline 126 & $\tau'$\\
\hline 234 & $\tau'$\\
\hline  135 & $\tau'^6+3\tau'^4+11\tau'^2+10$\\
\hline 136 & $\tau'(\tau'^2+4)$\\
\hline  235 & $\tau'(\tau'^2+4)$\\
\hline  145 & $\tau'(\tau'^6+\tau'^4+\tau'^2+2)$\\
\hline 146 & $\tau'(\tau'^4+3\tau'^2+5)$\\
\hline 245 & $\tau'(\tau'^4+3\tau'^2+5)$\\
\hline 156 & $\tau'^4$\\
\hline 345 & $\tau'^4$\\
\hline 236& $\tau'(\tau'^4+5\tau'^2+3)$\\
\hline 246& $\tau'(\tau'^6+3\tau'^4+11\tau'^2+10)$\\
\hline 256 & $\tau'^4(\tau'^2+4)$\\
\hline 346 & $\tau'^4(\tau'^2+4)$\\
\hline  356& $\tau'^4(\tau'^4+2\tau'^2+6)$\\
\hline  456& $\tau'^9$\\\hline

\end{tabular}

\caption{Evaluation of the punctured-disc polynomials  for $k=2,\
r=3$, $p=1$ and $N=2,3,4,5,6$. When $N$ is odd, the natural
variable is $\tau'^2$.} \label{tab3}
\end{table}

\clearpage

\section{Transfer matrix  \label{section.The transfer matrix}}

\subsection{$\Psi$ as an eigenvector of the transfer matrix}

In the disc case, at $t^3=q=1$, the vector $\Psi$
(\ref{psidefinition}) is an eigenstate of a commuting family
transfer matrix $T(z_0)=T(z_0|z_i)$ \cite{Pdf}. The matrix
elements of $T(z_0)$ are polynomials in the variables $z_0,z_i$
called the spectral parameters. We give a simple proof that in the
limiting case where cyclicity is recovered: $z_{i+N}=z_i$, a
vector $\Psi(z_i)$ obeying the duality relations (\ref{duality})
can be obtained as the stationary state of a transfer matrix.

In this case, by specializing the spectral parameters of the
transfer matrix, one obtains a Hamiltonian, $H=\sum_1^N e_i$, with
positive integer matrix elements having $\Psi(1,\dots,1)$ as a
stationary state (Notice that the cyclic operator $\sigma$ must be
defined in order to construct $e_N$). Thus, by the
Perron-Frobenius theorem, the components of $\Psi(1,\dots,1)$ can
be normalized to be positive integers. These integers are
conjectured to count ASM \cite{razumov1}. At this value, the sum
of the components of $\Psi(z_i)$ is a symmetric polynomial in the
$z_i$ which can be evaluated explicitly \cite{Pdf}. We give a
derivation of this property relying on the fact that the T.L.
algebra becomes non-semisimple and can be realized trivially,
($e_i=1$), at the corresponding values of $t$.

Let us define the following permutation operators acting on the
spins at positions $i$ and $j$:
\begin{eqnarray}
 Y_{ij}({z_i\over z_j})={z_jT_{ij}-z_iT_{ij}^{-1}\over z_it^{1\over 2}-z_jt^{-{1\over 2}}}.
 \label{Ydef2}
\end{eqnarray}

\bigskip
The vector $\Psi$ can also be characterized by the conditions
equivalent to (\ref{duality})\cite{pas1}:
\begin{eqnarray}
 Y_i({z_i\over z_{i+1}})\Psi(\dots,z_i,z_{i+1}\dots)&=&\Psi (\dots,z_{i+1},z_{i}\dots)\cr
 \sigma\Psi(z_1,\dots,z_{N-1},z_N)&=&\Psi (z_2,\dots,z_N,q^{-1}z_{1}).
 \label{intertwine-psi}
\end{eqnarray}
The first relation (\ref{duality}) can be straightforwardly
verified by substituting (\ref{Ydef2}) into
(\ref{intertwine-psi}). Note that he normalization of $Y_{ij}$ has
been chosen so that $Y_{ii+1}e_i=e_i$, and therefore $\bar e_i$
(\ref{polynomeTLei}) projects onto a polynomial symmetric under
the exchange of $z_i,z_{i+1}$.

It is convenient to introduce the permutation $P_{ij}$ which act
by permuting the indices $i$ and $j$. One then defines the
operators $X_{ij}=P_{ij}Y_{ij}({z_i\over z_j})$ obeying the
Yang-Baxter equation (equivalent to the fact that the $Y_{ij}$ are
permutation operators):
\begin{eqnarray}
 X_{ij}X_{ji}&=&1, \cr
 X_{ij}X_{kl}&=&X_{kl}X_{ij},\ \ {\rm if}\ i,j\ne k,l,\cr
 X_{ij}X_{ik}X_{jk}&=&X_{jk}X_{ik}X_{ij},
\label{Yang-Baxter}
\end{eqnarray}

Using the Yang Baxter operator, we can define the commuting
operators $\tilde y_i$ by:
\begin{eqnarray}
 \tilde y_i&=&X_{ii-1}\dots X_{i1}\hat q_i\dots X_{ii+1} ,\label{yang-rep01}
 \label{qKZ}
\end{eqnarray}
with $\hat q_i$ acting from the right shifts the variable $z_i$ by
a factor $q$ (\ref{polyaction-si}). By commuting $\hat q_i$ to the
left, we replace the operators $X_{ij}$ located to its left by
$X_{ij+n}$. When we act from the left with $\tilde y_i$ on $\Psi$,
using repeatedly the first equation (\ref{intertwine-psi}), one
sees that the product of $X_{ik}$ substitutes the variables $z_i$
to $z_{i+n}$ in $\Psi$. Conversely, the operator $\hat q_i$
replaces $z_i$ with $z_{i+n}$. So $\Psi$ is an eigenvector with
eigenvalue $1$ of $\tilde y_i$:
\begin{eqnarray}
\tilde y_i\Psi=\Psi.
 \label{qKZ1}
\end{eqnarray}
In what follows, we restrict to the case where $\hat q_i=1$, and
we identify the transfer matrix as the generating function of the
$\tilde y_i$. The Transfer matrix has the expression:
\begin{eqnarray}
T(z_0|z_i)=\mathop{\rm {t r }}_0\{ X_{01}({z_0\over
z_1})X_{02}({z_0\over z_2})\dots X_{0n}({z_0\over
 z_N})\},
 \label{Tdef2}
\end{eqnarray}
where the partial trace (defined in section
\ref{section.string-pattern}) is on the label $0$.

It follows from (\ref{Yang-Baxter}) that two matrices with
different spectral parameters $z_0$ and $w_0$ and all the other
spectral parameters equal commute with each other
\cite{baxter}\cite{gaudin}:
\begin{eqnarray}
[T(z_0|z_i),T(w_0|z_i)]=0. \label{commute}
\end{eqnarray}

When the shift operator $\hat q_i=1$ which occurs when $q=1$, it
is straightforward to verify that $\tilde y_i$ is obtained by
substituting $z_i$ to $z_0$ in the expression of
$T(z_0)=T(z_0|z_j)$ \cite{gaudin}:
\begin{eqnarray}
\tilde y_i=T(z_i). \label{y=T}
\end{eqnarray}


\bigskip

In the pattern representation, and in the case $\tau=\tau'=u=1$,
the T.L. matrices $e_i$ (\ref{lsei}) transform a pattern into a
single pattern with a coefficient equal to one. It follows from
this property, that the line-vector having all its entries equal
to one is a left eigenvector of $T(z_0)$ with the eigenvalue $1$.
As a result, $T(z_0)$ has a right eigenvector with the eigenvalue
$1$ which we determine to be $\Psi$.

From (\ref{y=T}), $(T(z_0)-1)\Psi$ is a rational fraction, with a
numerator of degree $N$ in $z_0$. It vanishes when $z_0=z_i$. One
also has $T(0)=T(\infty)=1,$ and thus $(T(z_0)-1)\Psi$ also
vanishes when $z_0=0,\infty$. It is therefore equal to zero and
$\Psi$ is an eigenvector of $T(z_0)$ with the eigenvalue $1$.

\subsubsection{Sum of the components\label{section.cyclic}}

At the cyclic point $q=1$, it follows from the explicit expression
(\ref{lsei}) of the matrices $e_i$, that the vector $\chi$ with
all its entries equal to one is a left eigenvector of the $e_i$'s
with the eigenvalue one. Thus, the scalar product $\chi.\Psi$,
equal to the sum of the components of $\Psi$, is by the duality
relation (\ref{duality}) a symmetric polynomial. We obtain its
expression here (see also \cite{zuber}).

Up to the normalization factor, this sum is determined to be the
lowest degree {\it symmetrical} polynomial obeying the wheel
condition.

For the minimal degree disc representation generated by
(\ref{Fomega.disc0.def}) with $k=2$, the polynomial vanishes when
three ordered variables are specialized by the $r=2$ wheel:
$z_i=z,z_j=tz,z_k=t^2z$ for $i<j<k$. It has the degree:
\begin{eqnarray}
 \lambda=({N}-1,{N}-1,{N}-2,{N}-2,\dots,0,0),
 \label{degre}
\end{eqnarray}
if the number of variables is even $(N_1=N_2=N)$,  and the first
$N-1$ is erased if the number of variables is odd  $(N_1=N,\
N_2=N-1)$. This determines it to be equal to the Schur function
$s_{\lambda}$ \cite{pas1}.

An analogous discussion can be made in the punctured disk case.
$Z_N$ can be determined by using a recursion argument on $N$ as
follows. Let us normalize $Z_N(z_i)$ so that its highest degree
monomial $Z_N(z_i)$ is equal to $z_1^{N-1}z_2^{N-2}\dots z_N^0$.
Using the projection (\ref{definitE'}), one has:
\begin{eqnarray}
E'(Z_N)(z_3,\dots )= Z_{N-2}(z_3,\dots ). \label{recursion}
\end{eqnarray}
By recursion, this condition determines $Z_N(z_i)$ completely to
be the product of two Schur functions:
\begin{eqnarray}
Z_N(z_i)=S_{\lambda_1,N}S_{\lambda_2,N},
\end{eqnarray}
where $\lambda_1=(0,0,1,1,2,2\dots )$ and
$\lambda_2=(0,1,1,2,2\dots )$ and each partition has $N$ rows.
Indeed, the product $S_{\lambda_1,N}S_{\lambda_2,N}$ has the same
highest degree monomial as $Z_N$, and each Schur function
factorizes as follows when one specializes the values of the two
first variables:
\begin{eqnarray}
 S_{\lambda_1,n}(z_1=z,z_2=tz,z_3,\dots )&=&\prod_{i=3}^N (t^2z-z_i)
 S_{\lambda_1,N-2}(z_3,\dots )\cr
S_{\lambda_2,n}(z_1=z,z_2=tz,z_3,\dots )&=&z \prod_{i=3}^N
(t^2z-z_i)
 S_{\lambda_2,N-2}(z_3,\dots )
 . \label{recursion2}
\end{eqnarray}
The product has the same highest monomial as $Z_N$ and obeys the
same recursion relation (\ref{recursion}), it is therefore equal
to $Z_N$.

\section{Spin and SOS representations\label{section.sos.rep}}

We characterize some representation of the T.L. affine algebra
$\mathcal{A}^T_N(t)$ acting on on a spin or a path basis.  The
representation depends on two positive integer $r$ and $s$, and on
a continuous parameter $u$. When $t$ is generic, and for $|r-s|\le
1$, these representations are isomorphic to the patten
representations of section (\ref{section.link-pattern}).

In the spin basis, $r$ and $s$ are respectively the number of $+$
and $-$ spins, $u$ is a twist parameter, and the operators $y_i$
are realized as triangular matrices. On the other hand, the paths
form an orthogonal basis and the affine generators $y_i$ are
realized as diagonal matrices in this basis.

\subsection{Paths\label{section.paths}}

We describe here the action of the generators of the A.H.A. on the
paths of the SOS representation. This essentially the
specialization to the two column tableaux of the construction done
in the appendix \ref{appendice-Permutation operators}. The spin
basis also admits a representation described in the appendix
(\ref{section.diag-yi-spin}). In the generic case, the two basis
are related by a triangular transformation, and the affine
generators $y_i$ are realized as triangular matrices in the spin
basis and as diagonal matrices in the path basis. There is
analogous of the K-L basis in the spin case known as the canonical
basis \cite{lus2} which we do not discuss here.

The path basis states are directed paths $\pi$ on the square
lattice. The path $\pi$ is a sequence of lattice points
$\pi_i=(x_i^+,x_i^-),\ 0\le i\le N $. It starts from the origin
$(0,0)$, and moves by steps of one unit towards the north-east or
the south-east:
\begin{eqnarray}
(x_{i+1}^+,x_{i+1}^-)=(x_{i}^++1,x_{i}^-)\ {\rm or} \
(x_{i}^+,x_{i}^-+1),
 \label{motion-path}
\end{eqnarray}
to reach the final point $(x_N^+,x_N^-)=(N_+,N_-)$. The path can
also be described by $(i,h_i)=(x_i^+ +x_i^-,x_i^+ -x_i^-)$ where
$h_i$ defines the height of the path point $i$. Thus the path
starts from the height $0$ to reach the height $N_+-N_-$ in $N_+$
up and $N_-$ down steps.

The affine generator  $y_i$ acts diagonally on a path by looking
at the $i^{\rm th}$ step between $i-1$ and $i$. It is equal to
$\hat y_+(x_{i-1}^+)$ if this step is up (towards the north-east),
and to  $\hat y_-(x_{i-1}^-)$ if it is down (towards the
south-east), where $\hat y_+(a)=y_+t^{-a}$ and $\hat
y_-(a)=y_-t^{-a}$.

Let us determine the expression of the T.L. generators so that the
relations (\ref{T.L.},\ref{affine-gener}) are satisfied. Since
$e_i$ commutes with $y_l$ for $l\ne i,i+1$, we require that $e_i$
acts locally on the piece of path between $i-1$ and $i+1$.

Therefore, the projector $e_i$ is equal to zero if $h_{i-1}\ne
h_{i+1}$, and it decomposes into block matrices, $e_i=\oplus\
\delta_{h_{i-1}-h} e^h$, where $e^h$ is equal to zero on paths
such that $h_{i-1}\ne h$ or $h_{i+1}\ne h$. It acts as a two by
two matrix on a pair of paths equal everywhere except at the three
consecutive points $(i-1,i,i+1)$ where their heights take the
values: $[h,h-1,h]\ {\rm and}\ [h,h+1,h].$

Then, writing the last relation of (\ref{affine-gener}) in terms
of the T.L. generators:
\begin{eqnarray}
t^{-{1\over 2}}y_i-t^{1\over 2}y_{i+1}=y_{i+1}e_{i}-e_{i}y_{i},
 \label{eq.ei.path.equ}
\end{eqnarray}
and substituting the matrix of $e^{h}$ into this equality, we
determine its diagonal elements. Finally, by requiring that $e_i$
is proportional to a projector and satisfies (\ref{T.L.}), we
determine $e^h$ up to a diagonal similarity transformation to be
given by \cite{Pasthese}:
\begin{eqnarray}
e^h=-{1\over S_h}\pmatrix{S_{h-1}&S_{h+1}\cr S_{h-1}&S_{h+1}},
 \label{ei.path.expr}
\end{eqnarray}
where $S_h$ is defined as:
\begin{eqnarray}
S_h=y_-t^{h\over 2}-y_+t^{-{h\over 2}},
 \label{S.expr}
\end{eqnarray}
and obeys the recursion relation:
\begin{eqnarray}
S_{h-1}+S_{h+1}=-\tau S_h.
 \label{S.recurrence}
\end{eqnarray}
With this normalization, the paths form an orthogonal basis and
the square of their norm is $\prod_i S_i^{{}}$.

In the appendix \ref{appendice-Permutation operators}, we define
operators $Y_l$ which permute the two paths $h_{l}=h\pm 1$ when
$h_{l-1}=h_{l+1}=h$.

\bigskip

The two possible eigenvalues $y_\pm$ of $y_1$ are defined up to a
common factor. It is convenient to fix this normalization by
setting:
\begin{eqnarray}
 y_+=ut^{N_+-1\over 2},\ y_-=u^{-1}t^{N_--1\over 2},
  \label{ur-us}
 \end{eqnarray}
so that we have:
\begin{eqnarray}
 y_1y_2\dots y_N=\sigma^N=u^{N_+-N_-}.
  \label{normal-yi}
 \end{eqnarray}

We can identify this representation with the spin representation
of the appendix (\ref{section.diag-yi-spin}) with the same value
of $u$ by taking $N_+$ to be the number of $+$ spins and $N_-$ to
be the number of $-$ spins.

We also recognize the pattern representations of section
(\ref{section.link-pattern}) if we  identify the values of $u$,
and take:
\begin{eqnarray}
 N_+&=&{N\over 2},\ N_-={N\over 2},\ {\rm if\ }N\ {\rm \ is\ even},\nonumber \\ \cr
 N_+&=&{N+1\over 2},\ N_-={N-1\over 2},\ {\rm if\ }N\ {\rm \ is\ odd}.
 \label{comparaison-spin}
 \end{eqnarray}

The spin representation coincides with the pattern representation
with $N_+=N_-={N\over 2}$ if $N$ is even and $N_+={N+1\over 2},$
$N_-={N-1\over 2}$  if $N$ is odd.

\subsubsection{Restricted paths}

If parameterize $u$ as $u=t^{{k\over 2}}$ and define the spin of
the path to be $S^z={N_+-N_-\over 2}$, we can characterize the
representation by its basis states given by paths of $N$ steps
starting from the initial height $h_0=-k-S^z$ and reaching the
final height $h_N=-k+S^z$. We denote $\rho^{(N)}_{hh'}$, the
representations obtained for $h_0=h, h_N=h'$.

The factors $y_+,y_-$ in the definition of $S_h$ are absorbed in
the redefinition of the height so that one has:
\begin{eqnarray}
S_h=t^{h\over 2}-t^{-{h\over 2}},
 \label{S.expr2}
\end{eqnarray}
and:
\begin{eqnarray}
 y_+=t^{-{h\over 2}}t^{N-1\over 4},\ y_-=t^{{h'\over 2}}t^{N-1\over
 4}.
 \label{S.expr2}
\end{eqnarray}
\bigskip

If $h_0$ is integer, for $S_h$ to be defined, the height must be
restricted to be strictly positive.

We can encode a path into a two lines tableau where the numbers
$i=1,2,\dots ,n$ are successively registered in the first or the
second line according to whether the  $i^{\rm th}$ step is towards
the north-east or the south-east. The abscissa of the first line
are $h_0,h_0+1,\dots$ and the abscissa of the second line are
$1,2,\dots$.

For $h_0$ generic, one obtains in this way the standard tableau
with two lines of length $N_+$ and $N_-$ described in the appendix
\ref{appendice-Permutation operators}.



\section{Coxeter-Dynkin diagram representations and action
of the modular group on the trace\label{section.rsos-paths}}

This section lies somewhat outside the scope of the paper. We
consider more specifically the representations $\rho^N_{hh'}$ of
the affine T.L. algebra $\mathcal{A}^T_N(t)$, in the root of unity
case. We construct representations associated to a Coxeter-Dynkin
diagram which we decompose into the $\rho^N_{hh'}$. This
decomposition is {independent}  of $N$. It is consistent with an
action of the modular group, which leaves the Coxeter-Dynkin
diagram trace invariant but acts on $\rho^N_{hh'}$. It can be
viewed as a finite size version of the modular invariant partition
function of conformal the field theories (CFT).\footnote{A
geometrical interpretation of the splitting into two independent
numbers $h,h'$ can be given from the three dimensional Topological
Quantum Field Theory (or annular tensor category) point of view
\cite{walker}.}

\bigskip

If the heights defining the paths of section \ref{section.paths}
are integer, we can restrict the paths to have a strictly positive
height. Similarly, when $t$ is a root of unity, $S_p=0$ in
(\ref{S.expr2}) and we can restrict the height to be strictly less
than $p$. The paths obeying these restrictions are called
restricted solid on solid (RSOS) paths \cite{baxter}.

In the RSOS case, the basis states of $\rho^{(n)}_{hh'}$ are the
paths of length $N$ starting from the height $h$ and ending at the
height $h'$, such that the the heights $h_i $ obey the constraint
$1\le h_i\le p-1$.

\bigskip

We now construct a representation of $\mathcal{A}^T_N(t)$
associated to an arbitrary finite bipartite graph ${\cal D}$ which
we identify with its incidence matrix \cite{Pasthese}. The Hilbert
$H^N_{\cal D}$ is defined by its orthogonal basis given by the
closed paths of length $N$: $|a_0,a_1,\dots,a_N=a_0\rangle$, drawn
on ${\cal D}$, such that the two vertices $a_i$ and $a_{i+1}$ are
adjacent on the graph.

Let $S_a$ be the components of an eigenvector of ${\cal D}$, and
$\tau=-(t+t^{-1})$ be the corresponding eigenvalue. The T.L.
generators $e_l$ are defined similarly as in (\ref{ei.path.expr}):
$e_l$ acts locally on the piece of path between $l-1$ and $l+1$,
it is equal to zero if $a_{l-1}\ne a_{l+1}$. It decomposes into
block matrices $e_{l}=\oplus \ e^a$ where $e^a$ acts on the pieces
of path: $|a_{l-1}=a,a_l=b,a_{l+1}=a\rangle=|b\rangle$ and is
given by:
\begin{eqnarray}
e^a_{bc}=-{S_{c}\over S_a}.
 \label{ei.path.expr.gen}
\end{eqnarray}
The cyclic operator $\sigma$ cyclically shifts the paths by one
unit: $\sigma|a_i\rangle=|a_{i-1}\rangle$.

These representations are particularly interesting in the case
where the incidence matrix of the diagram has a Perron-Frobenius
eigenvalue less than two. The diagram is then a Coxeter-Dynkin
diagram ${\cal D}\in A_m,D_m,E_{6},E_7,E_8,$ and we choose $S_h$
to be the Perron-Frobenius eigenvector of ${\cal D}$. $t$ is then
a primitive root of unity, $t=e^{2i\pi\over p}$, where $p$ is the
Coxeter-number of the diagrams given below:
\begin{eqnarray}
A_m,D_m,E_{6},E_7,E_8\leftrightarrow m+1,2(m-1),12,18,30.
 \label{coxeter}
\end{eqnarray}
We can decompose the Hilbert space $H^N_{\cal D}$ into the
irreducible representations $\rho^N_{hh'}$:
\begin{eqnarray}
 H^N_{\cal D}=\oplus\gamma^{\cal D}_{hh'} \rho^N_{hh'},
  \label{hilbert-decomposition}
\end{eqnarray}
where $\gamma^{\cal D}_{hh'}\in {\mathbb N}$ are the
multiplicities.

By adapting the arguments of \cite{pas-partition,saleur}, we
compute the multiplicities in the appendix \ref{appendix.Bi-Matrix
formalism} and we show that they are independent of $N$ when it is
large enough. The coefficients of the decomposition coincide with
the coefficients of the character decomposition of parafermionic
C.F.T. unitary models \cite{capelli}\cite{gepner}. More precisely,
one can define an action of the modular group which leaves the
trace of the ${\cal D}$ representations invariant and transforms
the trace of $\rho_{hh'}$ as the tensor product of the two
characters $\chi_h$ and $\chi_{h'}$ of the Virasoro algebra
entering the decomposition of the partition function.

\bigskip

\section{Conclusion}

We have deformed  the stationary state of a $O(1)$ model defined
on the disc and the cylinder. This has led us to study polynomial
representations of the affine Hecke algebra depending on two
complex parameters $t$ and $q$ related by the relation
$t^{k+1}q^{r-1}=1$. These polynomials obey some vanishing
conditions and interpolate between the stationary state of a
stochastic transfer matrix at $t^{k+1}=1$ and a Q.H.E. wave
function at $t=1$. In the cases presented here ($k=2$), the
transfer matrix is a that of a $O(n)$ model with $n=1$. Another
family not considered here interpolates between the stationary
state of a $O(1)$ model related to the Birman-Wenzl-Murakami
algebra and the Pfaffian state of the Q.H.E. \cite{pas2}.

One can distinguish a basis labelled by link patterns which in the
disc case coincides with the Kazhdan-Lusztig basis. We conjecture
that the specialization of the basis polynomials at $z_i=1$ are
{\it positive} in the loop fugacity. Moreover, when the loop
fugacity is equal to one, they count certain classes of
alternating sign matrices.

It is intriguing  that the spectral parameters of a transfer
matrix can also be viewed as the coordinates of particles moving
in the plane, so that the braiding properties of the polynomials
are analogous to the fractional statistics properties of the
Q.H.E. wave functions \cite{frank}. The vanishing conditions
obeyed by the polynomials are the q-deformed vanishing conditions
of the Q.H.E. wave functions \cite{moore}, and we suspect that
their positivity properties are in some way related to the
incompressible (minimal area for the quantum state which translate
into minimal degree condition) properties of the corresponding
Q.H.E. wave functions.

Finally, in another direction, we have  established that the
unitary representations of the A.H.A. at roots of unity obey
modular properties similar to those of conformal field theories.

\subsection{Acknowledgments}

We thank P. Di Francesco, T. Miwa, V. Jones, B. Nienhuis, A.V.
Razumov, Y.G. Stroganov, T. Suzuki, K. Walker, P. Zinn-Justin and
J-B. Zuber for interesting discussions and comments. We thank A.
Lascoux for explanations which helped us to considerably improve
the revised version of this paper.

This work is partially supported by Grant-in-Aid for JSPS Fellows
No. 17-2106 and by the ANR program ``GIMP", ANR-05-BLAN-0029-01.

\subsection{Note added in proof}

Since we proposed the positivity conjectures, some progress has
been made confirming their validity. P. Di Francesco has
conjectured that the sum of the evaluations at $z_i=1$ of the disc
polynomials (table \ref{tab1}) are related to a $q$-enumeration of
Totally Symmetric Self-Complementary Plane Partitions \cite{Pdf3}.
In collaboration with A. Lascoux, one of us has established a
similar relation for the highest degree coefficients in $\tau$ of
these evaluations \cite{pas-lasc}.

\bigskip

\appendix

\section{Affine algebras \label{section.hecke-algebra}}

In this appendix, we give the defining relations of the double
A.H.A.

The Hecke algebra depending on the parameter $t$,
$\mathcal{A}_N(t)$ (or $\mathcal{A}_N$ when there is no possible
confusion), is generated by the generators $T_1,T_2,\dots
,T_{N-1}$,\footnote{Sometimes, we use the notation $T_{ii+1}$ for
$T_i$.}  obeying the braid relations:
\begin{eqnarray}
  T_{i}T_{i+1}T_{i}&=& T_{i+1}T_{i}T_{i+1}\cr
  T_{i}T_{j}&=&T_{j}T_i\ \
 {\rm if} \ |i-j|>1.
 \label{braid}
 \end{eqnarray}
and the quadratic relation:
\begin{eqnarray}
(T_{i}-t^{1\over 2})(T_{i}+ t^{-{1\over 2}})=0,\label{hecke}
\end{eqnarray}
for $1\le i\le n-1$.

If we define the generators $e_i=T_{i}-t^{1\over 2}$, the $e_i$
are projectors obeying the Hecke relations:
\begin{eqnarray}
e_{i}^2&=&\tau e_{i}, \cr e_ie_j&=&e_je_i\ \ {\rm if} \ |i-j|>1\cr
e_{i}e_{i+1}e_{i}-e_{i}&=&
e_{i+1}e_{i}e_{i+1}-e_{i+1},\label{hecke1}
\end{eqnarray}
where ${ \tau=-t^{1\over 2}-t^{-{1\over 2}}}$.

The A.H.A. \cite{lusztig}, is an extension of the Hecke algebra
(\ref{hecke}) by the generators $y_i,\ 1\le i\le n$ obeying the
following relations:
\begin{eqnarray}
y_iy_j&=&y_jy_i\cr  T_iy_j&=&y_jT_i \ \ \ {\rm if}\ j\ne i,i+1 \cr
 T_{i}y_{i+1}&=&y_{i}T_{i}^{-1}\ \ {\rm if} \ i\le
N-1.\label{affine-gener}
\end{eqnarray}

The double A.H.A. \cite{BGHP,cherednick2}, is the extension of the
A.H.A. obtained by adjoining to it operators denoted $z_i$. The
$z_i$ obey the same commutation relations (\ref{affine-gener}) as
the affine generators $y _i$ with the generators $T_k$. It depends
on an additional parameter $q$.

\begin{eqnarray}
z_iz_j&=&z_jz_i\cr  T_iz_j&=&z_jT_i \ \ \ {\rm if}\ j\ne i,i+1 \cr
 T_{i}z_{i+1}&=&z_{i}T_{i}^{-1}\ \ {\rm if} \ i\le
N-1.\label{affine-gener-z}
\end{eqnarray}

The relations obeyed by the $z_i$ and the $y_i$ due to Cherednik
\cite{cherednick2} are given by:

\begin{eqnarray}
 y_1z_2y_1^{-1}z_2^{-1}&=&T_1^2\cr
y_i (\prod_{j=1}^N z_j)&=& q (\prod_{j=1}^N z_j) y_i, \cr
 z_i(\prod_{j=1}^N y_j)&=& q^{-1} (\prod_{j=1}^N y_j) z_i.
\label{Cherd}\end{eqnarray}

A more elementary presentation \cite{BGHP} is in term of the
$T_i$, $z_i$ and of a cyclic generator $\sigma$ defined as:
\begin{eqnarray}
\sigma=  T_{N-1}^{-1}...T_{1}^{-1}y_1 .\label{cycle}
\end{eqnarray}

We extend the definition of the variables $z_i,\ i\in {\mathbb
Z},$ by cyclicity:

\begin{eqnarray}
z_{i+N}=q^{-1} z_i.\label{def-q1}
\end{eqnarray}

Similarly, one can define a braid generator $T_N$ by $T_N=\sigma
T_{1}\sigma^{-1} $. Using the braid relations again, one gets
$\sigma T_N=T_{N-1}\sigma $, and one can extend the definition of
$T_i$ to $i\in {\mathbb Z}$: $T_{i+N}=T_i$.

The defining relations of $\sigma$ are then:

\begin{eqnarray}
 \sigma T_i=T_{i-1}\sigma, \cr
 \sigma z_i=z_{i-1}\sigma.
\label{affine-rep1-cycle}
\end{eqnarray}

The double affine Hecke algebra is generated by the generators
$T_i,z_i$ and  $\sigma$ (\ref{affine-rep1-cycle}). In the appendix
\ref{section.diag-yi}, we reconstruct the generators $y_i$ from
this presentation

\bigskip

In this paper, we are often concerned with the T.L. quotient
$\mathcal{A}_N^T(t)$ of the A.H.A. generated by $e_i,y_i$ where we
constrain the generators $e_i$ to obey the restrictions:
\begin{eqnarray}
e_{i}e_{i\pm 1}e_{i}&=&e_{i}.\label{T.L.}
\end{eqnarray}

\bigskip
\subsection{Intertwining operators}
\subsubsection{Tableaux representations of the A.H.A. \label{appendice-Permutation
operators}}

Following \cite{knop}, one can define operators $Y_{l}$ which
intertwine the affine generators:
\begin{eqnarray}
  y_j Y_{l}&=&Y_{l}  y_j \ {\rm if} \ j\ne l,l+1 \cr
  y_lY_{l}&=&Y_{l}  y_{l+1} \cr
  y_{l+1}Y_{l}&=&Y_{l}  y_l.
  \label{intertwine-Y}
\end{eqnarray}
The relations (\ref{intertwine-Y}) are satisfied by
\begin{eqnarray}
 Y'_l&=&y_lT_l^{-1}-y_{l+1}T_{l} ,\cr
&=&(t^{1\over 2}y_l-t^{-{1\over
2}}y_{l+1})+(y_l-y_{l+1})(\tau-e_l).
  \label{intertwine-Y^2}
\end{eqnarray}

Thus, $Y'_l$ intertwines the eigenstates of $y_l$ and $y_{l+1}$.
The square of $Y'_l$ acts diagonally on such states:
\begin{eqnarray}
Y_l'^2=(t^{{1\over 2}}y_{l}-t^{-{1\over2}}y_{l+1})(t^{{1\over
2}}y_{l+1}-t^{-{1\over2}}y_{l}),
  \label{intertwine-Y^2}
\end{eqnarray}
and is null on the states with $y_{l+1}=t^{\pm 1}y_l$.

\bigskip

We can use this property to construct a representation of the
Hecke algebra on standard tableaux. A standard tableau is a right
eigenvector of the $y_i$'s and the eigenvalue of $y_i$ depends on
the box of the diagram where the number $i$ is located. Each box
is assigned an eigenvalue so that a vertical move of one unit
north or a horizontal move of one unit east multiplies the
eigenvalue by a factor $t$.

Thus, the two eigenvalues of $y_l$ and $y_{l+1}$ differ by a
factor of $t$ when the numbers $l$ and $l+1$ belong to adjacent
boxes of the the same column with $y_l=ty_{l+1},$ or of the same
line with $y_{l+1}=ty_l$.
The tableau is annihilated by $Y'_l$ in both cases. It is
annihilated by $e_l$ in the column case and by $e_l-\tau$ in the
line case.

If $l$ and $l+1$ do not belong to adjacent boxes, $Y'_l$ exchanges
the position of $l$ and $l+1$. We can multiply $Y'_l$ by a
normalization factor equal to $(t^{{1\over
2}}y_{l}-t^{-{1\over2}}y_{l+1})^{-1}$ so that the square of the
resulting operator, $Y_l,$ is equal to one:
\begin{eqnarray}
 Y_{l}
  &=&1+{y_l-y_{l+1}\over t^{{1\over
  2}}y_{l}-t^{-{1\over2}}y_{l+1}}(\tau-e_l),.
 \label{Y.def}
\end{eqnarray}
If we adopt a normalization so that $Y_l$ permutes the two
tableaux, we recover the representation \cite{wenzl} of the Hecke
algebra on tableaux. Consider two tableaux which are exchanged by
$Y_l$. We denote $\hat y$ and $\hat y'$ the eigenvalue of $y_l$
and $y_{l+1}$ on the first tableau (and of $y_{l+1}$ and $y_{l}$
on the second tableau). The expression of $T_l$ in the basis made
by these two tableaux is given by:
\begin{eqnarray}
 T_l={1\over \hat y-\hat y'}\pmatrix{(t^{1\over 2}-t^{-{1\over 2}})\hat y&-t^{1\over 2}
 \hat y+t^{-{1\over 2}}\hat y'\cr
 -t^{-{1\over 2}} \hat y+t^{1\over 2} \hat y'&(t^{-{1\over 2}}-t^{1\over 2})\hat y'}.
 \label{e-tau}
\end{eqnarray}

\bigskip

In the path basis of section \ref{section.paths}, the paths are
annihilated by $Y_l$ if $h_{l+1}\ne h_{l-1}$. $Y_l$ acts locally
on the piece of path between $l-1$ and $l+1$ as follows: it is
equal to zero if $h_{l-1}\ne h_{l+1}$, and it decomposes into
block matrices $Y_{l}=\oplus \ Y_l^h$ where $Y_l^h$ acts on paths
with $h_{l-1}=h_{l+1}=h$ by exchanging the two paths with the
intermediate heights $h_l=h-1$ and $h_l=h+1$.
All the paths are obtained from one of them under the repeated
action of the generators $Y_l.$

In the cases where $h_l$ are integers, the paths must be
restricted to have $h_l>0$. Furthermore, when $t^p=1$, the paths
must be restricted to have $h_l<p$.






\subsubsection{Shift operators \label{section.Permutation and shift
operators.}}

Following \cite{knop}, we give the expression of an operator which
changes the degree of a polynomial by keeping it an eigenstate of
the operators $y_i$.

As in (\ref{def-q1}) for the coordinates $z_i$, it is convenient
to extend the definition of the affine generators $y_i$ to $i\in
{\mathbb Z}$ by periodicity:
\begin{eqnarray}
 y_{i+N}=q^{-1} y_i.
 \label{def-yi.periodique}
\end{eqnarray}

Using the affine Hecke relations (\ref{affine-gener-z}), one can
construct a shift operator $\bar A$ which shifts by one unit the
affine generators:
\begin{eqnarray}
 \bar A\bar y_i&=&\bar y_{i+1}\bar A\cr
 \bar A\bar g_i&=&\bar g_{i+1}\bar A
 \label{shift-operator-polynome}
\end{eqnarray}
It raises the degree of the polynomial by one and its expression
is given by
\begin{eqnarray}
 \bar A&=&z_1\bar\sigma^{-1}.
\label{shift-oper-polynome-expr}
\end{eqnarray}
\bigskip

\section{Explicit matrices in the cases n=2,3\label{appendix.one line}}


In the basis $\alpha\beta,\beta\alpha$, the action of
$\mathcal{A}^T_2$ is generated by the matrices:
\begin{eqnarray}
e_1=\pmatrix{\tau&\tau'\cr 0&0}, \ \sigma =\pmatrix{0&1\cr 1&0}.
\label{TL4}
\end{eqnarray}

The T.L. algebra relation (\ref{T.L.}) is modified into:
$e_1e_2e_1=\tau'^2e_1$ because the left hand side creates two
lines around the torus in this case.

\bigskip

In the basis  $\alpha\alpha\beta$, $\alpha\beta\alpha,$
$\beta\alpha\alpha$ the generators of $\mathcal{A}^T_3$ are given
by:

\begin{eqnarray}
 e_1=\pmatrix{0&0&0\cr 1&\tau&u\cr 0&0&0}, \
 \sigma=\pmatrix{0&0&1\cr u&0&0\cr 0&1&0}.
 \label{birman-matrix-H4}
 \end{eqnarray}

 The matrix $y_1$ is given by:
\begin{eqnarray}
 y_1=\pmatrix{u&u^{-1}&-t^{-{1\over 2}}\cr0&u^{-1}t^{-{1\over 2}}+u&-t^{-1}\cr
 0&t^{1\over 2}&0}.
\label{birman-matrix-H4-2}
\end{eqnarray}

\section{\bf Representation of the  $\bar y_j$ in the polynomial
and the spin cases.\label{section.diag-yi}}

\subsection{polynomial representation\label{section-polynomial
representation}}

We repeat here the diagonalization of the operators $\bar y_j$
done in \cite{BGHP}. The method follows the one initiated by
Sutherland \cite{Sutherland} in the Calogero-Sutherland model
context.

For a certain ordering of the monomial basis, we obtain an
expression of $y_i$ in a triangular form by decomposing it into a
product of triangular matrices $x_{ij}$ and a diagonal operator
$\hat q_i$.


The expression of $\bar y_i$ which follows from
(\ref{cycle}) and (\ref{affine-gener}) is given by:
\begin{eqnarray} \bar y_i&=&
\bar T_i\bar T_{i+1}\dots \bar T_{N-1}\bar \sigma \bar
T_1^{-1}\dots \bar T_{i-1}^{-1}.
\label{cycle2}\label{expression.yi}
\end{eqnarray}

It is convenient to reexpress $y_i$ as a product of triangular
operators. for this, we decompose the cyclic operator sigma as a
product of elementary permutation $s_i$ which permute the
variables $z_i$ and $z_{i+1}$ and an operator $\hat q_1$ which
replace the variable $z_1$ with $q^{-1}z_1$:

\begin{eqnarray}
\bar \sigma=c_0 s_{N-1}\dots s_{2}s_{1}\hat q_1.\label{sigma-k}
\end{eqnarray}
with $\hat q_i$ given by:

\begin{eqnarray}
 P(z_1,\dots,z_i\dots,z_N)\hat q_{i}&=&
P(z_1\dots ,q^{-1}z_i,\dots ,z_N).\label{polyaction-si}
\end{eqnarray}

 We define $x_{ij}=\bar
T_{ij}s_{ij}$ where $s_{ij}$ is the permutation operator which
permutes the variables $z_i$ and $z_j$. $\bar T_{ij}$ is defined
in the same way as $\bar T_{i}=\bar T_{ii+1}=t^{1\over 2}+\bar
e_i$ where we replace the variable $z_{i+1}$ with the variable
$z_j$ and the permutation $s_i$ with permutation $s_{ij}$ in the
expression (\ref{polynomeTLei}) of $\bar e_i$ . The operator
$x_{ij}$ takes the form for $i<j$:
\begin{eqnarray}
x_{ij}=-t^{-{1\over 2}}+(t^{{1\over 2}}-t^{-{1\over
2}})(1-s_{ij}){z_j\over z_i-z_j},\label{polyactionX}
\end{eqnarray}

After permuting the $s_i$ through, the expression of $\bar y_i$
becomes:
\begin{eqnarray}
\bar y_i&=&{c_0 x_{ii+1}x_{ii+2}\dots x_{iN}\hat q_i
x_{1i}^{-1}\dots x_{i-1i}^{-1}}\label{yang-rep00}
\end{eqnarray}

The operator $x_{ij}$ commutes with $z_iz_j$ and acts as a
triangular matrix on the monomials $z_i^m,\ z_j^m $:
\begin{eqnarray}
z_i^mx_{ij}&=&-t^{-{1\over 2}}z_i^m+(t^{{1\over 2}}-t^{-{1\over
2}})(z_i^{m-1}z_j+z_i^{m-2}z_j^2+\dots +z_j^m)\ \ {\rm if}\ m\ge
0,\cr z_j^mx_{ij}&=&-t^{{1\over 2}}z_j^m-(t^{{1\over
2}}-t^{-{1\over 2}})(z_i^{m-1}z_j+z_i^{m-2}z_j^2+\dots
+z_iz_j^{m-1})\ {\rm if}\ m> 0 .\label{ordreX}
\end{eqnarray}

Thus, the operators $\bar y_i$ are also realized as triangular
operators in the monomial basis.

We order on the monomials by saying that
$z^{\lambda}=z_1^{\lambda_1}\dots z_1^{\lambda_N}$ is larger than
$z^{\mu}=z_1^{\mu_1}\dots z_N^{\mu_N}$ if either $\mu$ is obtained
from $\lambda$ by a sequence of squeezing operations
$\{\lambda'_i,\lambda'_j\}\to \{\lambda'_i-1,\lambda'_j+1\}$ with
${\lambda'_i>\lambda'_j+1},$ or $\mu$ is a permutation of
$\lambda$ and can be obtained from $\lambda$ by a sequence of
permutations $(\lambda'_i,\lambda'_{i+1})\to
(\lambda'_{i+1},\lambda'_{i})$ with $\lambda'_i>\lambda'_{i+1}$ .

It follows from the expressions (\ref{ordreX}) that the action of
$y_j$ on a monomial produces only monomials which are smaller with
respect to this order, and their eigenvalues are given by the
diagonal elements in the monomial basis.

A common  eigenstate of the operators $\bar y_i$ is characterized
by its highest degree monomial $z^{\lambda_\pi}$ where $\pi$ is
the shortest permutation such that
$(\lambda_\pi)_i=\lambda^+_{\pi_i}$ and $\lambda^+$ is a partition
($\lambda^+_1\ge \lambda^+_2\dots \ge\lambda^+_N$).

The action of $\bar y_j$ on a monomial $z^{\lambda_\pi}$ is given
by
\begin{eqnarray}
z^{\lambda_\pi}\bar y_j=c_0(-t^{-{1\over 2}})^{N-1}
q^{-\lambda^+_{\pi_j}}t^{\pi_j-1}z^{\lambda_\pi}+{\rm lower\
terms} ,\label{spectre-yj}
\end{eqnarray}
from which the eigenvalues of $\bar y_j$ follow.

The global normalization of $\bar y_i$ is such that:
\begin{eqnarray}
 \bar y_1\bar y_2\dots \bar y_N=\bar \sigma^N=c_0^N\hat q_1\dots\hat q_{N}=c_0^Nq^{-|\lambda|}.
 \label{normalization-yj}
\end{eqnarray}

\subsection{Conditional expectation value in the T.L. cases
\label{section-Conditional expectation value in the T.L}}

When the A.H.A. reduces to a T.L. algebra $\mathcal{A}^T_N$, in
the case of minimal degree polynomials with $k=2$, we can define a
projection $E'$ from the polynomials in $N$ variables to the
polynomials in $N-2$ variables dual to the inclusion defined in
(\ref{project-2}).

For any polynomial $F$, $E'$ satisfies the conditions:
\begin{eqnarray}
 &a)& \ \ E'( F e_1)=\tau E'(F)\cr
 &b)& \ \ E'( F x)=E'( F)x\ \forall x\in \mathcal{A}^T_{N-2}\cr
 &c)&\ \ E'(F e_1)=0\Rightarrow F e_1=0,
 \label{condition1E'}
\end{eqnarray}
and  can be realized  as:
\begin{eqnarray}
E'(F)(z_3,z_4,\dots )=\phi(z,z_3,z_4,\dots
)^{-1}F(z_1={z},z_2=tz,z_3,z_4,\dots ), \label{definitE'}
\end{eqnarray}
where $\phi(z,z_i)$ is a polynomial which removes the $z$
dependence of $F$ in the right hand side of (\ref{definitE'})
equal to:
\begin{eqnarray}
\phi(z,z_i)&=&z^p\prod_{i=3}^{N}\prod_{b=0}^{r-2}(t^2q^bz-z_i).
\label{definitE'2}
\end{eqnarray}

\subsection{Spin representation\label{section.diag-yi-spin}}

The spin $1/2$ representation of ${\mathcal A}_N$ can be obtained
from the representation on polynomials with a degree less or equal
to $1$ in each variable. The monomials $z^{\lambda}$ are the spin
basis elements: $\lambda_i=1$ if the spin $i$ is plus and $0$ if
it is minus.

Let us describe this representation explicitly. The Hilbert space
is $({\mathbb{C}^2})^N$ with a basis given by sequences of spins
$|\pm\pm\dots\pm\rangle$. The matrix $e_i$ acts in
$\mathbb{C}^2_i\otimes\mathbb{C}^2_{i+1}$ and has the following
expression in the basis
$|++\rangle,|+-\rangle,|-+\rangle,|--\rangle:$
\begin{eqnarray}
 e_i=\pmatrix{0&0&0&0\cr 0&-{t^{-{1\over 2}}}&1&0\cr 0&1&-t^{{1\over 2}}&0\cr 0&0&0&0}.
 \label{spin-matrix-e}
 \end{eqnarray}

The permutation operators $P_{ij}$ permute the spins at positions
$i$ and $j$. It is convenient to introduce the operators $e_{ij}$
having the expression (\ref{spin-matrix-e}) and acting in
$\mathbb{C}^2_i\otimes\mathbb{C}^2_{j}$. $T_{ij}=t^{1\over
2}+e_{ij}$, and the operators $x_{ij}$ are defined as in
(\ref{section-polynomial representation}), $x_{ij}=T_{ij}P_{ij}.$
In the same basis as above, $x_{ij}$ is realized as a triangular
matrices as:

\begin{eqnarray}
 x_{ij}=\pmatrix{t^{{1\over 2}}&0&0&0\cr 0&1&t^{{1\over 2}}-{t^{-{1\over 2}}}&0\cr 0&0&1&0\cr 0&0&0&{t^{{1\over 2}}}}.
 \label{spin-matrix-x}
 \end{eqnarray}

The diagonal matrix $\Omega_i$ acts on the spin at position $i$.
In the basis $|+\rangle,|-\rangle$, it is given by:
\begin{eqnarray}
 \Omega_i=\pmatrix{u^{}&0\cr 0&u^{-1}}.
 \label{spin-matrix-om}
 \end{eqnarray}

The matrix $\sigma$ is defined by (\ref{sigma-k}) where we
substitute  $\Omega_i$ to $\hat q_i$ and $P_{ii+1}$ to $s_{ii+1}$:
\begin{eqnarray}
\sigma=P_{N-1N}\dots P_{12}\Omega_1.
 \label{spin-matrix-sigma}
 \end{eqnarray}
It implies the following identification of spins:
$\mu_{i+N}=u^{-\mu_i}\mu_i$.

The Hilbert space is characterized by the total numbers $N_+,\
N_-$, of plus and minus spins respectively $(N_++N_-=N)$. The
representation also depends on the parameter $u$ of the twist
matrix $\Omega$.

We can give an alternative definition of the spin representation
with the total spin $N_+-N_-\over 2$. Let
$\mathcal{A}^T_{(N_+,N_-)}$ be a subalgebra of $\mathcal{A}^T_N$
generated by
 $T_1,\ldots,T_{N_+-1},T_{N_++1},\ldots,T_{N-1}$, and $y_1,\ldots,y_N$.
We define a one-dimensional representation $\mathbb{C}{\bf1}$ of
$\mathcal{A}_{(N_+,N_-)}$ as follows:
\begin{eqnarray}
T_i {\bf1}&=& t^{1\over2}{\bf1}, \\
y_i {\bf1}&=&ut^{N_+-1 \over 2}t^{i-1} {\bf1} \quad\mbox{if $1\leq i\leq N_+$}, \\
y_i {\bf1}&=&u^{-1}t^{N_--1 \over 2}t^{i-N_+-1} {\bf1}
\quad\mbox{if $N_++1\leq i\leq N$}.
\end{eqnarray}
Then the spin representation is isomorphic to an induced module
$\mathrm{Ind}_{\mathcal{A}_{(N_+,N_-)}}^{\mathcal{A}_N}
\mathbb{C}{\bf1}$, where $|++\cdots+--\cdots-\rangle$ corresponds
to ${\bf1}$.

We can define an ordering on the basis as follows. A state
$|\mu'\rangle$ is smaller than $|\mu\rangle$, if it can be
obtained from $|\mu\rangle$ through a sequence of permutations of
a plus spin at position $i$ and a minus spin at position $i+1$.


\section{Stability of the wheel condition under the action of the
A.H.A. \label{section.proof stability}}

We show here that the wheel conditions are preserved under the
action of the A.H.A. generators.

\bigskip

Let us verify  that the ideal of polynomials obeying the condition
(\ref{wheelcondition2}) under the restriction
(\ref{wheelcondition1}a) is preserved under the action of $\bar
e_i$ (\ref{polynomeTLei}).

Consider the polynomial $P'=P\bar e_{i}$ where $P(z_i)$ is a
polynomial obeying any admissible wheel condition. We show that
$P'$ obeys the wheel condition specified by any admissible wheel
$\{i_a\}$ and $\{b_{aa+1}\}$.

\hskip .5cm If there exists a value $\bar a$, such that $i=i_{\bar
a}$, $i+1=i_{\bar a+1}$, and $b_{\bar a \bar a+1}=0$, then $\tau
P-P'$ is proportional to $t^{1\over 2}z_i-t^{-{1\over 2}}z_{i+1}$
and obeys this wheel condition. By linearity, so does $P'$.

\hskip .5cm If not, the wheel deduced by permutation: $i'_{
a}=s_{i}(i_a)$ and $b'_{aa+1}=b_{aa+1}\ \forall a$ is admissible.
Thus, $Ps_i,$ and by linearity $P'$, also obey this wheel
condition.

\bigskip

Let us show that the conditions (\ref{wheelcondition0}) and
(\ref{wheelcondition1}b) imply that the space of polynomials
obeying the wheel conditions is preserved under the action of
$\bar\sigma$ (\ref{sigma-definition}). This amounts to show that
the transformation defined by ${i_a}\to {i_a+1}$ with
$\{b_{aa+1}\}$ unchanged defines an admissible wheel condition.
The identification $z_{N+1}=q^{-1}z_1$ (\ref{def-q}), is used when
$i_a+1=N+1$.
\smallskip

\hskip .5cm If $i_a<N \ \forall a,$ it is obvious.

\hskip .5cm If $i_{\bar a}=N$ for a value $\bar a\ne k+1$, then
$i_{\bar a+1}<N$ and $b_{\bar a\bar a+1}\ge 1$. The above
transformation can be recast into the wheel condition:
$i'_{a}=i_a+1$ for $a\ne \bar a$, $i'_{\bar a}=1$, and $b'_{\bar
a-1\bar a}=b_{\bar a-1\bar a}+1$, $b'_{\bar a\bar a+1}=b_{\bar
a\bar a+1}-1$, $b'_{aa+1}=b'_{aa+1}$ for $a$ and $a+1\ne \bar a$.

\hskip .5cm If $i_{k+1}=N$, this transformation can be recast into
the wheel condition:  $i'_{a}=i_{a-1}+1$ for $a\ge 2$, $i'_1=1$,
and $b'_{aa+1}=b_{a-1a}$ for $a\ge 2,$ $b_{12}=r-2-\sum_1^k
b_{aa+1}$. To express $b_{12}$ in this form we have used the
condition (\ref{wheelcondition0}), and the condition
(\ref{wheelcondition1}b) is necessary to have $b_{12}\ge 0$. We
also have: $\sum_1^k b'_{aa+1}\le r-2$ and the condition
(\ref{wheelcondition1}b) is satisfied by the new wheel.

\section{Partition functions and traces\label{appendix.Bi-Matrix
formalism}}

In this appendix, we give a graphical method to evaluate the trace
of operators in $\mathcal{A}^T_N(t)$ following
\cite{pas-partition}. The trace depends on the representation. We
compare the traces between the different representations, and
obtain the decomposition of $\rho_{\cal D}$ into the
representations $\rho_{hh'}$, by showing the trace identity:

\begin{eqnarray}
 {\rm t r}_{\cal D}(x)=\sum_{hh'}\gamma^{\cal D}_{hh'}
 {\rm t r}_{\rho_{hh'}}(x),\forall x\in \mathcal{A}^T_N(t).
  \label{hilbert-decomposition-trace}
\end{eqnarray}
In this identity, the matrices are finite dimensional. We fix $t$
to be the root of unity, $t=e^{2i\pi\over p}$, where $p$ is the
Coxeter number of the diagram ${\cal D}$.

We also define an action of the modular group which leaves ${\rm t
r}_{\cal D}$ invariant but transforms linearly the traces ${\rm t
r}_{\rho_{hh'}}$. The coefficients $\gamma^{\cal D}_{hh'}$
therefore define a modular invariant decomposition of the trace.

\bigskip
To represent the trace, we use the description of linear operators
by a system of lines drawn on the annulus of section
\ref{section.link-pattern}. We close the annulus into a torus by
identifying the two boundaries.  The trace becomes the partition
function of a loop model on the torus.

The contractible loops can be removed by giving them a weight
$\tau$. One ends up with a system of non-contractible loops not
touching each other and therefore homotopic to the same cycle.

\smallskip

In the case of the models defined by a diagram ${\cal D}$ of
section \ref{section.rsos-paths}, if the number loops is $2m$, the
weight is the number of paths of length $2m$ which can be drawn on
the diagram ${\cal D}$.

In the case of the spin representation of the appendix
\ref{section.diag-yi-spin}, the lines carry a spin index and are
oriented accordingly. We must then sum over the possible
orientations of the loops. We cut the annulus into a rectangle as
in figure \ref{fig:affine}.
If the total spin across a horizontal cycle is $2{S^z}$, and the
total spin across a vertical cycle is $2{S^z}'$. The weight of an
oriented loop configuration is $u^{2{S^z}'}$.
\bigskip

We  denote by ${k \choose S^z}$ the spin-representation of
$\mathcal{A}^T_N(t)$ where the value of the spin is fixed to be
$S^z$ and $u=t^{{k\over 2}}$. If $|S^z|>{N\over2}$, we regard ${k
\choose S^z}$ as the zero representation.

We define two actions $s_0$ and $s_1$ on a 2-tuple ${a \choose b}$
by:
\begin{eqnarray}
s_1{a \choose b}&=&{b \choose a}\\
s_0{a \choose b}&=&{b-p \choose a+p}.
\end{eqnarray}

Setting $h=-k-S^z$ and $h'=-k+S^z$, we conjecture that:
\begin{eqnarray}
{\rm tr}_{\rho_{hh'}}(x)=\sum_{w\in W} (-1)^{l(w)}{\rm tr}_{w{k
\choose S^z}}(x) \quad,\forall x\in \mathcal{A}^T_N(t)
  \label{hilbert-decomposition-trace-rhof}
\end{eqnarray}
where $W=W(A_1^{(1)})$ is the affine Weyl group of type
$A_1^{(1)}$. Note that the right hand side is a finite sum.

To establish the identity (\ref{hilbert-decomposition-trace}), it
is useful to introduce an intermediate representation $\rho_f$
defined by a graph ${\cal D}_f$ made of $2f$ vertices connected
around a circle. We require that $f$ divides $p$ so that
$\tau=-(t+t^{-1})$ is an eigenvalue of ${\cal D}_f$. We obtain a
representation of $\mathcal{A}^T_N(t)$ if we define the T.L.
generators $e_l$ as in (\ref{ei.path.expr.gen}) by taking for
$S_a$ the eigenvector of ${\cal D}_f$ with the eigenvalue $\tau$.
We can also view the trace of this representation as the partition
function of a spin model where the total spins $2S^z$ and
${2S^z}'$ across the horizontal and vertical cycles considered
above are constrained to be equal to $f$ modulo $p$. We set
$p=ff'$. By representing the constraint on ${2S^z}'$ as a Fourier
sum, we obtain the decomposition:
\begin{eqnarray}
 \rho_f= 2\oplus {af'\choose b f}\ {\rm with}\ 0\le a\le f-1,\ 0\le b\le f'-1.
  \label{hilbert-decomposition-trace-rhof-1}
\end{eqnarray}

We can decompose the representations $\rho_{\cal D}$ in terms of
the $\rho_f$ by identifying their traces as in
(\ref{hilbert-decomposition-trace}). With this interpretation we
write $\rho_{\cal D}=\sum c_f \rho_f$ where $f$ is a divisor of
$p$. From the property of the trace, we determine the coefficients
$c_f$ by requiring that the number of closed paths of a given
length on the graph ${\cal D}$ is the same as the sum over $f$ of
the number of closed paths of the same length on the circular
diagram ${\cal D}_f$ multiplied by $c_f$. One obtains
\cite{pas-partition}:
\begin{eqnarray}
 2\rho_{{ A}_n}&=&\rho_{n+1}-\rho_{1},\cr
 2\rho_{{ D}_n}&=&\rho_{2(n-1)}-\rho_{n-1}+\rho_{2}-\rho_{1},\cr
 2\rho_{{ E}_6}&=&\rho_{12}-\rho_{6}-\rho_{4}+\rho_{3}+\rho_{2}-\rho_{1},\cr
 2\rho_{{ E}_7}&=&\rho_{18}-\rho_{9}-\rho_{6}+\rho_{3}+\rho_{2}-\rho_{1},\cr
 2\rho_{{E}_8}&=&\rho_{30}-\rho_{15}-\rho_{10}-\rho_{6}+\rho_{5}+\rho_{3}+\rho_{2}-\rho_{1}.
 \label{hilbert-decomposition-trace-rhof-2}
\end{eqnarray}
Finally, by combining (\ref{hilbert-decomposition-trace-rhof},
\ref{hilbert-decomposition-trace-rhof-1} and
\ref{hilbert-decomposition-trace-rhof-2}), we obtain the following
decomposition:
\begin{eqnarray}
 \rho_{{ A}_n}&=&\oplus_{k=1}^{n}\rho_{k,k},\cr
 \rho_{{D}_{2n}}&=&\oplus_{k=1}^{n-1}\rho_{(2k-1)+(4n-2k-1),(2k-1)+(4n-2k-1)}\oplus
 2\rho_{2n-1,2n-1},\cr
 \rho_{{D}_{2n+1}}&=&\oplus_{k=1}^{2n}\rho_{2k-1,2k-1}\oplus\rho_{2n,2n}\oplus_{k=1}^{n-1}(\rho_{2k,4n-k}
 \oplus\rho_{4n-k,2k}),\cr
 \rho_{E_6}&=& \rho_{1+7,1+7}\oplus \rho_{4+8,4+8}\oplus \rho_{5+11,5+11},\cr
 \rho_{E_7}&=&\rho_{1+17,1+17}\oplus \rho_{5+13,5+13}\oplus \rho_{7+11,7+11}
 \oplus \rho_{9,9}\oplus \rho_{5+13,9}\oplus \rho_{9,5+13},\cr
 \rho_{E_8}&=&\rho_{1+11+19+29,1+11+19+29}\oplus\rho_{7+13+17+23,7+13+17+23}.
 \label{hilbert-decomposition-3}
\end{eqnarray}
where
$\rho_{a+b,c+d}=\rho_{a,c}\oplus\rho_{a,d}\oplus\rho_{b,c}\oplus\rho_{b,d}$.

Let us obtain the transformation law of the trace of $\rho_{hh'}$
under a modular transformation. The spin across the vertical and
horizontal cycles is transformed as:
\begin{eqnarray}
 {{S^z}'\choose S^z}\to \pmatrix{a&b\cr c&d}{{S^z}'\choose S^z}.
 \label{modular-1}
\end{eqnarray}
From the characterization preceding
(\ref{hilbert-decomposition-trace-rhof}) of $\rho_{hh'}$, it is
straightforward to obtain the following transformation of the
traces:
\begin{eqnarray}
 \pmatrix{1&1\cr 0&1}\ &:&\ {\rm t r}_{\rho_{hh'}}\to t^{{1\over 4}({h'^2-h^2
 })}{\rm t r}_{\rho_{hh'}},\cr
 \pmatrix{1&0\cr 1&1}\ &:&\ {\rm t r}_{\rho_{hh'}}\to
 \sum_{rr'}t^{{1\over 4}((r-h)^2-(r'-h')^2)}{\rm t
 r}_{\rho_{rr'}}.
 \label{modular-2}
\end{eqnarray}
 Notice that under a modular transformation, the
representation $\rho_{hh'}$ behaves as a tensor product of affine
characters of $A^{(1)}_1$ : $\rho_{hh'}\sim \chi_l \otimes
\bar\chi_{l'}$, where the level $k$ is given by $p=2(k+2)$ and the
spin $l$ is given by $h=2l+1$.
\bigskip

Under a modular transformation of the torus, the partition
function of the ${\cal D}$-models remains invariant but the
partition functions of $\rho_{hh'}$ transforms linearly.
Therefore, the multiplicities $\gamma_{hh'}$ in
(\ref{hilbert-decomposition-3}) are such that the direct sum is
left invariant under these transformations.


\begin{thebibliography}{99}

\bibitem{razumov1} A.V. Razumov and Y.G. Stroganov, {\it Spin chains and combinatorics,}
J.Phys. A {\bf 34}, 3185,
[cond-mat/0012141].

\bibitem{pierce} P.A. Pearce, V. Rittenberg, J. de Gier and B. Nienhuis,
{\it Temperley-Lieb Stochastic Processes} J.Phys.A {\bf 35}
L661-668 (2002) [math-phys/0209017].


\bibitem{batchelor} M.T. Batchelor, J. de Gier and B.Nienhuis,
J.Phys.A {\bf 34} L265-270 (2001) [cond-mat/0101385].

\bibitem{razumov2} A.V. Razumov and Y.G. Stroganov,
{\it Combinatorial nature of ground state vector of $O(1)$ loop
models} Theor.Math.Phys. {\bf 138} 333-337 (2004).


\bibitem{Pdf} P. Di Francesco and P. Zinn-Justin,
{\it Around the Razumov-Stroganov conjecture: proof of a
multiparameter sum rule,} Electr.J.Combin. {\bf12}, R6 (2005),
[math-ph/0410061].

\bibitem{kasatani} M. Kasatani, {\it Subrepresentations in the polynomial representation of
the double affine Hecke algebra of type $GL_n$ at
$t^{k+1}q^{r-1}=1$,} Int. Math. Res. Not. {\bf 2005}, no. 28,
1717--1742 [math. QA/0501272].


\bibitem{pas1} V. Pasquier, {\it Quantum incompressibility and
Razumov stroganov type conjectures}, Ann. Henri Poincar\'e {\bf 7}
397-421 (2006) [cond-mat/0506075].


\bibitem{cherednick2} I. Cherednik, {\it Double Affine Hecke Algebras.}
Cambridge University Press (2005).

\bibitem{BGHP} D. Bernard, M. Gaudin, D. Haldane and V. Pasquier,
{\it Yang-Baxter equation in spin chains with long range
interactions},  J.Phys. A {\bf 26}, 5219-5236 (1993),
[hep-th/9301084].


\bibitem{feigin} B. Feigin, M. Jimbo, T. Miwa, E. Mukhin, {\it
Symmetric polynomials vanishing on the shifted diagonal and
Macdonald polynomials}, Int. Math. Res. Not. {\bf 18}, 101 (2003).

\bibitem{Macdo1} I.G. Macdonald, {\it A new class of symmetric
functions.} Actes {\bf$20^e$} Seminaire Lotharingien, p 131-171,
Publications I.R.M.A. Strasbourg (1988), 372/S-20.


\bibitem{macdo2}I.G. Macdonald, {\it Symmetric Functions and Hall
Polynomials.}, Oxford University Press, (1995).



\bibitem{read}  E. Prange and S. Girvin, {\it The Quantum Hall
effect.} Springer-Verlag, (1987).




\bibitem{Haldane-Rezayi} F.D.M. Haldane and E.H. Rezayi, Phys.
{\it Spin-singlet wave function for the half-integral quantum Hall
effect, } Rev. Lett. {\bf 60}, 956, and {\bf E60}, 1886 (1988).

\bibitem{KL} D. Kazhdan and G. Lusztig, {\it Representation of
Coxeter groups and Hecke algebras,} Invent.Math. {\bf 53}, 165-184
(1979).


\bibitem{lasc1} A. Lascoux, M.-P. Sch\"utzenberger,
{\it Polyn\^omes de Kazhdan $\&$ Lusztig pour les grassmaniennes},
Asterisque {\bf 87-88}, 249-266 (1981).

\bibitem{lasc2} A. Lascoux, M.-P. Sch\"utzenberger,
{\it Symmetry and Flag manifolds}, Invariant Theory, Springer L.N
{\bf 996}, 118-144 (1983).

\bibitem{lasc-kiril} A. Kirillov, Jr.
and A. Lascoux, {\it Factorization of Kazhdan-Lusztig elements for
Grassmanians} Adv. Stud. {\bf 28}  143-154 (2000), [math/9902072].


\bibitem{fulton} W. Fulton {\it
Young Tableaux,} London Mathematical Society Student Texts {\bf
35} (2003) Cambridge University Press.

\bibitem{nienhuis} S. Mitra and B. Nienhuis, {\it Osculating Random Walks on Cylinders}
[math-ph/03120336].

\bibitem{nienhuis2} S. Mitra and B. Nienhuis, {\it Exact conjectured expressions for
correlations in the dense $O(1)$ loop model on cylinders }
[cond-mat/0407578].

\bibitem{razumov3} A.V. Razumov and Y.G. Stroganov,
{\it Enumeration of half-turn symmetric alternating sign matrices
of odd order,} [math-phys/0504022].




\bibitem{lusztig} G. Lusztig, {\it Affine Hecke algebras and their graded
version,} J.Amer.Math.Soc. {\bf 2}, 599-635 (1989).

\bibitem{lus2} G. Lusztig, {\it Introduction to Quantum Groups,} Birkhauser, Boston, (1993).

\bibitem{cherednick} I. Cherednik, {\it Nonsymmetric Macdonald Polynomials,}
Internat. Math. Res. Notices {\bf 10}, 483-515 (1995).



\bibitem{Pdf2} P. Di Francesco and P. Zinn-Justin,
{\it Inhomogenous model of crossing loops and multidegrees of some
algebraic varieties} [math-ph/0412031].

\bibitem{zuber} P. Di Francesco, P. Zinn-Justin and J.B. Zuber,
{\it Sum rules for the ground states of the O(1) loop model on a
cylinder and the XXZ spin chain,} [math-ph/0603009].


\bibitem{Jones} F.M. Goodman, P. de la Harpe, V.F.R. Jones, {\it Coxeter
Graphs and Towers of Algebras.} Springer-Verlag (1989).

\bibitem{lehrer} J.J. Graham and G.I. Lehrer, Enseign.Math. {\it The representation theory of
affine Temperley-Lieb algebras} {\bf 44} (1998).









\bibitem{Pasthese} V. Pasquier, {\it Two-dimentional critical systems
labelled by Dynkin diagrams.} Nucl.Phys.B  {\bf FS19}, 162-172
(1987).

\bibitem{pas-partition} V.Pasquier, {\it Lattice derivation of
modular invarient partition functions on the torus} J.Phys.A {\bf
20} L1229 (1987).


\bibitem{saleur} V. Pasquier and H. Saleur, {\it Common structures
between finite systems and conformald field theories through
quantum groups} Nucl.Phys.B  {\bf 330}, 525-556 (1990).


\bibitem{capelli} A. Cappelli, C. Itzykson and J.B. Zuber,
{\it Modular invariant partition functions in two dimensions}
Nucl.Phys.B  {\bf 280}[FS18], 445-465 (1987).

\bibitem{gepner} D.Gepner and Z.Qiu,
{\it Modular invariant partition functions for parafermionic field
theories} Nucl.Phys.B {\bf 285}[FS19], 423-453 (1987).

\bibitem{laughlin} E. Prange and S. Girvin, {\it The Quantum Hall
effect.} Springer-Verlag, (1987).

\bibitem{knop} F. Knop and S. Sahi,
{\it A recursion and a combinatorial formula for Jack polynomials,
}Invent. Math. {\bf 128} no.1, 9-22 (1997) [q-alg/9610016].

\bibitem{Pdf-hecke} P. Di Francesco and P. Zinn-Justin, {\it The quantum Knizhnik-Zamolodchikov
equation, generalized Razumov-Stroganov sum rules and extended
Joseph polynomials,} J.Phys.A {\bf 38} L815-822 (2002)
[math-ph/0508059].


\bibitem{ram} A. Ram {\it Affine Hecke algebras and generalized standard Young tableaux}, J. Algebra {\bf 260} (2003), no. 1, 367--415 [math.RT/0401329].

\bibitem{suzuki} T. Suzuki and M. Vazirani, {\it Tableaux on periodic skew diagrams
and irreducible representations of the double affine Hecke algebra
of type A}, Int. Math. Res. Not. {\bf 2005}, no. 27, 1621--1656
[Math.QA/0406617].



\bibitem{walker} K. Walker, private communication.


\bibitem{gaudin} M. Gaudin, {\it La Fonction d'Onde de Bethe,}
Masson (1981).


\bibitem{baxter} R.J. Baxter, {\it Exactly solved Models in Statistical Mechanics,} Academic, London (1982).

\bibitem{wenzl} H. Wenzl, {\it Hecke Algebras of type $A_n$ and subfactors,}
Invent.Math. {\bf 92}, 349-383 (1988).

\bibitem{affine} V. Pasquier, {\it Scattering Matrices and Affine Hecke
Algebras}, Schladming School 1995, Nucl.Phys.B (Proc.Suppl.) {\bf
45A}, 62-73(1996), [q-alg/9508002].


\bibitem{pas2} V. Pasquier, {\it Incompressible representations of the
Birman Wenzl algebra,}  Ann. Henri Poincar\'e {\bf 7} 603-619
(2006) [math.QA/0507364].

\bibitem{lecture} V. Pasquier, {\it A lecture on the Calogero Sutherland models},
{ The third Baltic Rim Student Seminar,} Saclay preprint,
Spht-94060 (1994), [hep-th/9405104].

\bibitem{hal-pas} F. D. M. Haldane, Z. N. C. Ha, J. C. Talstra, D. Bernard, and V.
Pasquier,
{\it Yangian symmetry of integrable quantum chains with long-range
interactions and a new description of states in conformal field
theory}, Phys. Rev. Lett. {\bf 69}, 2021–2025 (1992).



\bibitem{frank} F. Wilczek, {\it Fractional Statistics and Anyon
Superconductivity},
World Scientific.

\bibitem{moore} G. Moore and N. Read, {\it Nonabelions in the fractional quantum hall effect,}
, Nucl.Phys. B{\bf 360} 362-91 (1991).



\bibitem{Sutherland} B. Sutherland, Phys.Rev.A {\bf 5}, 1372
(1972).


\bibitem{Pdf3} P. Di Francesco, {\it Totally Symmetric Self-Complementary Plane Partitions
and Quantum Knizhnik-Zamolodchikov equation: a conjecture.}
[cond-mat/0607499].

\bibitem{pas-lasc} A. Lascoux and V. Pasquier, in preparation.

\
\end{thebibliography}
\end{document}